\documentclass[useAMS,usenatbib]{mn2e}

\usepackage{aas_macros,amsmath,graphicx,lineno,longtable,amssymb}

\newcommand{\rearth}{$R_{\oplus}$}
\newcommand{\mearth}{$M_{\oplus}$}
\newcommand{\water}{H$_2$O}
\newcommand{\kep}{{\it Kepler}}

\newcommand{\ha}{H${\alpha}$}
\newcommand{\mpy}{mas~yr$^{-1}$}
\newcommand{\teff}{$T_{\rm eff}$}
\newcommand{\tess}{TESS}
\newcommand{\mps}{m~sec$^{-1}$}
\newcommand{\tycho}{{\it Tycho}-2}

\title[Catalog Of Nearby Host Stars for Habitable
  Exoplanets]{Trumpeting M Dwarfs with CONCH-SHELL: a Catalog of
  Nearby Cool Host-Stars for Habitable ExopLanets and Life}

\author[Gaidos et al.]{E. Gaidos,$^{1}$\thanks{E-mail: gaidos@hawaii.edu (EG)} A.~W. Mann,$^{2}$\thanks{Harlan J. Smith Postdoctoral Fellow, University of Texas at Austin}, S. L\'{e}pine,$^{3,4}$, A. Buccino,$^{5,6}$, D. James,$^{7}$\newauthor
M. Ansdell,$^{2}$ R. Petrucci,$^{5}\thanks{Visiting Astronomer, Complejo Astron\'{o}mico El Leoncito, Argentina}$ P. Mauas,$^{5}$ and E.~J. Hilton$^{1,2}$\\
  $^{1}$Department of Geology \& Geophysics, University of Hawaii at M\={a}noa, Honolulu, Hawaii 96822 USA\\
  $^{2}$Institute for Astronomy, University of Hawaii at M\={a}noa, Honolulu, Hawaii 96822 USA\\
  $^{3}$Department of Physics \& Astronomy, Georgia State University, Atlanta, GA 3030 USA\\
  $^{4}$Department of Astrophysics, American Museum of Natural History, New York, NY 10024 USA\\
  $^{5}$Instituto de Astronom\'{i}a y F\'{i}sica del Espacio, C1428EHA - Buenos Aires, Argentina\\
$^{6}$Departamento de F\'{i}sica, FCEN-Universidad de Buenos Aires, Argentina\\
  $^{7}$Cerro Tololo Inter-American Observatory, Casilla 603, La Serena, Chile\\
}

\begin{document}
\date{Accepted to MNRAS}
\pagerange{\pageref{firstpage}--\pageref{lastpage}} \pubyear{2013}

\maketitle

\label{firstpage}

\begin{abstract}
We present an all-sky catalog of 2970 nearby ($d \lesssim 50$~pc),
bright ($J< 9$) M- or late K-type dwarf stars, 86\% of which have been
confirmed by spectroscopy.  This catalog will be useful for searches
for Earth-size and possibly Earth-like planets by future space-based
transit missions and ground-based infrared Doppler radial velocity
surveys.  Stars were selected from the SUPERBLINK proper motion
catalog according to absolute magnitudes, spectra, or a combination of
reduced proper motions and photometric colors.  From our spectra we
determined gravity-sensitive indices, and identified and removed 0.2\%
of these as interloping hotter or evolved stars.  Thirteen percent of
the stars exhibit \ha{} emission, an indication of stellar magnetic
activity and possible youth.  The mean metallicity is [Fe/H] = -0.07
with a standard deviation of 0.22~dex, similar to nearby solar-type
stars.  We determined stellar effective temperatures by least-squares
fitting of spectra to model predictions calibrated by fits to stars
with established bolometric temperatures, and estimated radii,
luminosities, and masses using empirical relations.  Six percent of
stars with images from integral field spectra are resolved doubles.
We inferred the planet population around M dwarfs using \kep{} data
and applied this to our catalog to predict detections by future
exoplanet surveys.
\end{abstract}


\section{Introduction}

The NASA \kep{} mission monitored approximately 200,000 stars for
transiting planets, and thousand of candidate planets have been
identified in the light curves \citep{Rowe2014}.  A few hundred of
these have been vetted and the overall rate of false positives is
generally, but not uniformly, low
\citep{Santerne2012,Colon2012,Fressin2013}.  Statistical analysis of
the candidates shows that at least half of stars host planets with
orbital periods less than $\sim$200~d and that Earth- to Neptune-size
planets are far more numerous than Jupiter-size planets
\citep[e.g.][]{Howard2012,Fressin2013}.  Some \kep{}-detected planets
orbiting very cool (late K- and M-type) dwarfs are near or inside the
theoretical ``habitable zones'' of these stars where an Earth-like
planet could have liquid water on its surface
\citep{Dressing2013,Kopparapu2013b,Gaidos2013,Quintana2014}.  But
\kep{} planet-hosting stars are typically distant (hundreds or
thousands of pc) and faint ($V \sim 15$), making measurement of mass
by Doppler radial velocity \citep[RV, e.g.][]{Marcy2014} or follow-up
such as transit spectroscopy or observations of secondary eclipse
difficult or impossible.

The \kep{} field covers only 0.25\% of the sky and, ironically, we
know much less about Earth- to Neptune-size planets around nearby
stars, including those around very cool dwarfs.  The RV method can be
readily applied to nearby stars which are widely spaced on the sky,
but because of the scaling between planet mass and radius, it is
comparatively less sensitive to smaller planets than the transit
method.  The most sensitive RV surveys have found a few super-Earths
on close orbits around M dwarfs \citep{Bonfils2013} but such surveys
have been hampered by the faintness of such stars at visible
wavelengths.  Ground-based, wide-field transit surveys are affected by
correlated (``red'') noise from the atmosphere and can only detect
short-period giant planets around F and G dwarfs.  Transit surveys of
nearby very cool dwarfs using individual pointings have met with
limited success \citep{Berta2013,Gaidos2014}.

Our knowledge of nearby small planets should dramatically improve with
two developments: the deployment of infrared Doppler spectrographs
that can exploit a spectral range where M dwarfs are brighter
\citep{Tamura2012,Thibault2012,Mahadevan2012,Quirrenbach2012}, and the launch of the
NASA Transiting Exoplanet Survey Satellite mission
\citep[TESS,][]{Ricker2010}.  In principle these surveys will detect
nearby Earth- or super-Earth-size planets on close-in orbits around
the brightest M dwarfs and/or measure their mass.  But until recently,
no all-sky catalog of well-characterized M dwarfs suitable as a source
of targets was available.  \citet[][hereafter LG11]{Lepine2011}
published a catalog of 8889 bright ($J< 10$), nearby late K- and early
M-type dwarfs selected from the SUPERBLINK proper motion catalog
\citep{Lepine2005} using proper motions and $V$-$J$ and $JHK_S$
colors.  This was followed by a spectroscopic survey of all the
northern LG11 stars with $J<9$ \citep{Lepine2013}.
\citet[][,hereafter F13]{Frith2013} also produced a catalog of bright
($K_S<9$) M dwarf candidates based on the PPMXL catalog
\citep{Roeser2010}.

Although these catalogs represent advances in cataloging and
describing the nearest M dwarf stars, there is still room for
improvement.  The spectroscopic catalog of \citet{Lepine2013} only
included stars at declinations $\delta > 0$.  The conservative
infrared color cuts imposed by LG11 to weed out giant stars also
eliminated some metal-rich dwarf stars which have red $J$-$H$ and
$H$-$K_S$ colors \citep{Leggett1992,Newton2014}.  These metal-rich
dwarfs are more likely to host giant planets
\citep{Johnson2009,Rojas-Ayala2010,Mann2013b}, e.g. HIP~79431
\citep{Apps2010}.  Finally, visible-wavelength ($BVg'r'i'$) photometry
is now available for most bright stars from the AAVSO Photometric All
Sky Survey \citep[APASS,][]{Henden2012}.  This CCD photometry is much
more precise ($\ge 0.02$ magnitudes) than the photographic plate-based
United States Naval Observatory \citep[USNO-B,]{Monet2003} magnitudes
used for most LG11 stars.  These data allow for more accurate
elimination of hotter or evolved stars based on photometric colors and
magnitudes.

For these reasons we have constructed a revised catalog, which we call
CONCH-SHELL (Catalog Of Nearby Cool Host-Stars for Habitable
ExopLanets and Life).  CONCH-SHELL is selected from the SUPERBLINK
catalog using modified criteria and new photometry and spectroscopy
(Section \ref{sec.catalog}).  Including previous data, we obtained
moderate-resolution ($\lambda/\Delta \lambda \sim 10^3$) spectra of
86\% of the catalog (Section \ref{sec.observations}).  We used spectra
to measure gravity-sensitive indicators, confirm the dwarf luminosity
class of these stars, and estimate their spectral type, effective
temperature, and metallicity (Section \ref{sec.parameters}).  For most
stars we measured any H$\alpha$ emission, an indicator of stellar
activity, and limited imaging of some stars allowed us to identify
binaries.  We combined effective temperatures with empirical relations
derived from observations of calibrator stars to estimate stellar
radius, luminosity, and mass.  We compared our catalog with F13
(Section \ref{sec.frith}).  We estimated the yield of transiting
planets that might be detected by \tess{} and future infrared Doppler
observations of these stars (Section \ref{sec.tess}).  We summarize
the properties of our catalog and the potential for future follow-up
observations in Section \ref{sec.summary}.

\section[]{Catalog Construction }
\label{sec.catalog}

LG11 selected candidate M dwarfs as stars that were (i) bright
($J<10$) (ii) red ($V-J > 2.7$), (iii) have absolute magnitudes or
reduced proper motions, proxies for absolute magnitudes, consistent
with the main sequence and (iv) infrared Two Micron All-Sky Survey
\citep[2MASS, ][]{Skrutskie2006} $JHK_S$ colors that are consistent with
M dwarfs.  \citet{Lepine2013} obtained spectra of the brightest
($J<9$) LG11 stars in the northern celestial hemisphere and showed
that they were virtually all M dwarfs.  In this work, we constructed a
revised catalog of $J<9$ M dwarfs using modified criteria and new
photometry from APASS.  Our criteria are based on a subset of stars
confirmed by either parallaxes or spectra from \citet{Lepine2013}.

We examined 21901 proper-motion stars in the SUPERBLINK catalog
\citep{Lepine2005} with $J<9$, $J$-$K_s>0.65$ and proper motion $\mu
>$ 40 \mpy{} (north of -20$^{\circ}$) or $\mu >80$ \mpy{} (south).
The north-south difference in proper motion limits reflects the higher
reliability of the SUPERBLINK catalog in the north.  Positions and
$JHK_s$ magnitudes and their errors were obtained by matching stars to
2MASS sources after correcting for proper motion over the difference
of the 2MASS observation epoch and 2000.  Our matching criterion was
based on the distribution of separations was 1 arc-sec and we flagged
stars where the angular separation is larger.  Six stars with
magnitudes flagged in the 2MASS catalog as being of poor quality or
upper limits due to detector nonlinearity are not used in computing
infrared colors (see below), but we do use the $J$ magnitudes in such
cases because an upper limit only means that the star is even brighter
and redder than stated.  Photographic visual magnitudes $V_E$ were
generated for all stars from USNO-B $b$ and $r$ magnitudes and 2MASS
$J$ magnitudes according to the prescription in \citet{Lepine2005}.

We matched selected SUPERBLINK stars to the revised version of the
{\it Hipparcos} catalog \citep{vanLeeuwen2007}, the \tycho{} catalog
\citep{Hog2000}, the APASS catalog (Date Release 7), and the All-Sky
Compiled Catalogue Version 2.5 \citep[ASCC-2.5, ][]{Kharchenko2009}.
The last catalog includes both {\it Hipparcos} and \tycho{} so there
is some redundancy.  The Hipparcos catalog was matched to the
SUPERBLINK stars assuming an observation epoch of 1992.25, calculated
by minimizing the median angular separation of matches, and differing
slightly from the nominal catalog epoch of 1991.25.  Based on the
distribution of matches, we applied a matching criterion of
$<1.3$~arcsec in angular separation and less than one magnitude
difference in Hipparcos $V$ vs. $V_E$.  Likewise, we used matching
criteria of 1.3~arcsec and 1.2 magnitudes for the \tycho{} catalog.
We found that the catalog epoch that minimized the median angular
separation was 1992.3, close to the {\it Hipparcos} epoch but much
earlier than the 2000 epoch given in the catalog's documentation.  For
APASS matches, we required that $0.8 < i-J < 2.8$ ($i$ from APASS and
$J$ from 2MASS) and an angular separation $<2.5$~arcsec.  If there was
more than one match to any star in the APASS catalog (25 cases) only
the closest match was considered, and we flagged these cases.

\tycho{} magnitudes were adjusted to the Johnson system using the
relationship in the Appendix of \citet{Mamajek2002}.  About 5\% of
stars have APASS $V < 10$ and we adjusted APASS $V$ magnitudes for
nonlinearity by comparing with {\it Hipparcos} photometry, calculating
a running median with a 0.25-magnitude bin, and fitting a line with
iterative 3$\sigma$ rejection of points.  The APASS-{\it Hipparcos}
offset at $V=11.2$ is only -0.093 magnitudes and the slope is 0.038
mag mag$^{-1}$.  We then calculated a color correction to convert
$V_E$ magnitudes to corrected APASS magnitudes by a linear fit to the
median difference vs. $V_E-J$ color in 0.2 magnitude bins.  The
APASS-$V_E$ offset at $V_E-J = 2.7$ is 0.081 magnitudes, and the
slope is -0.174 mag mag$^{-1}$.  The latter is apparently due to
imperfect calibration of $V_E$ against $V$ in \citet{Lepine2005}.
Visual magnitudes, as available, were assigned to stars in the
following order of decreasing priority: {\it Hipparcos}, APASS,
\tycho{}, and USNO-B.

We used parallaxes from the {\it Hipparcos} catalog as well as from
\citet{Harrington1993,vanAltena1995,McCook1999,Myers2002,Costa2005,Jao2005,Costa2006,Henry2006,Smart2007,Gatewood2008,Gatewood2009,Khrutskaya2010,Riedel2010,Jao2011}
and \citet{Dittmann2014}.  Some stars are proper motion companions to
{\it Hipparcos} stars and so have precise parallaxes but no {\it
  Hipparcos} numbers.

\begin{figure}
\includegraphics[width=84mm]{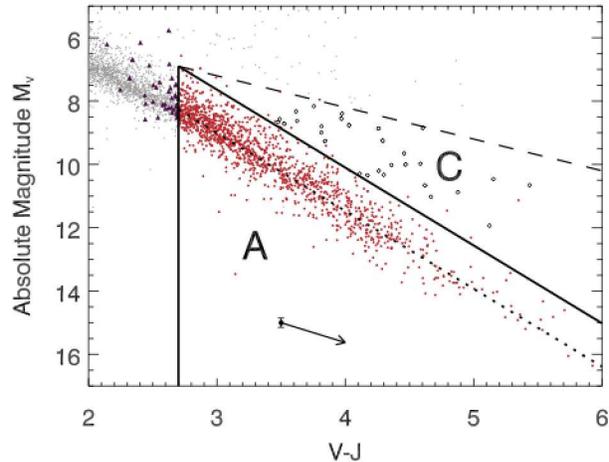}
\caption{Selection of M dwarfs based on absolute $V$-magnitude $M_V$
  vs. $V$-$J$ color.  Red points represent ``A-class'' stars that are
  spectroscopically confirmed M dwarfs or have $V-J > 2.7$ and $M_V$
  no brighter than three standard deviations of the main sequence
  locus (dotted line, see Section \ref{sec.catalog} for detailed
  selection criteria).  The upper solid curve is the main sequence
  locus plus three times the intrinsic width of the locus (0.46
  magnitudes).  The open points represent ``C-class'' M dwarfs that
  lie significantly above the main sequence but within a zone bounded
  by a slope one (dashed) line.  This zone could be populated by stars
  with large errors in $V$ or $J$, binaries or very young and
  relatively luminous stars.  Grey points represent other SUPERBLINK
  stars that were not selected.  The point with the arrow indicates
  the median error in $M_V$ and the direction of extinction.  Purple
  triangles are stars in the F13 catalog that were excluded from the
  CONCH-SHELL.}
 \label{fig.absmag}
\end{figure}

Absolute magnitudes $M_V$ were calculated for 9567 stars in the input
catalog with parallaxes. These are plotted vs. $V$-$J$ color in
Fig. \ref{fig.absmag}.  To describe the main sequence locus for $M_V$
vs. $V$-$J$ we iteratively fit a quadratic formula to median values in
a running 0.2 magnitude-wide bin with color.  The intrinsic scatter
(standard deviation) of the locus after accounting for errors in $M_V$
was re-computed for each iteration and only stars within three
standard deviations of the locus (where errors and intrinsic scatter
were added in quadrature) were retained for the next iteration.  The
final locus had an intrinsic width of 0.46 magnitudes, presumably due
to the metallicity dependence of luminosity and unresolved binaries.
We selected 1321 stars with $V$-$J>2.7$ and having $M_V$ within
$3\sigma$ of the final locus and more than $3\sigma$ fainter than $M_V
= 4.2$ (a threshold for identifying evolved stars) as M dwarfs.  To
these were added 622 stars with $V$-$J>2.7$ that were spectoscopically
confirmed as M dwarfs in \citet{Lepine2013}.  These 1943 stars are
plotted as the red points in Fig. \ref{fig.absmag}.

\begin{figure}
\includegraphics[width=84mm]{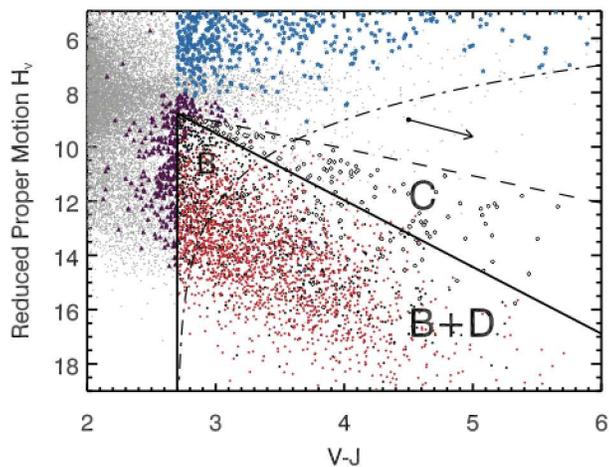}
\caption{Selection of M dwarfs based on reduced proper motion $H_V$
  vs. $V$-$J$ color.  The symbols and lines are the same as in
  Fig. \ref{fig.absmag}, with the addition that blue points are
  evolved/giant stars with $M_V < 4.2$.  ``B-class'' M dwarfs,
  represented by black points, have $V-J > 2.7$, $H_V$ fainter than
  the $M_V$ of the main sequence locus plus an offset of +0.5
  magnitudes (solid curve), and $g-r$ and $r-J$ colors consistent with
  M dwarfs.  ``C-class'' M dwarfs (open points) have $H_V$ below the
  dashed slope-one line and colors consistent with M dwarfs.
  ``D-class'' M dwarfs (also open points) have reduced proper motions
  consistent with M dwarfs but colors inconsistent with M dwarfs.  The
  dotted-dashed line approximately describes the maximum reddened
  $V-J$ that should be observed for an interloping star with
  unreddened $V$-$J$=2.7 and a given $H_V$.  D-class stars in a ``danger
  zone'' to the left of this line are excluded.}
 \label{fig.redpm}
\end{figure}

We identified additional M dwarfs lacking measured parallaxes based on
their reduced proper motions:
\begin{equation}
H_V = V + 5 \log \mu + 5.
\end{equation}
Stars with large proper motions, i.e. $H_V$ fainter than the main
sequence $M_V$ for their $V$-$J$ color, plus an offset, were selected
as M dwarfs (Fig. \ref{fig.redpm}).  We chose the offset to be 0.5
magnitudes based on an inspection of the distribution of $H_V - M_V$
values.  This criterion corresponds to a minimum transverse velocity
with respect to the Sun of 6~km~sec$^{-1}$.  The solar peculiar
velocity with respect to the Local Standard of Rest is itself about
18~km~sec$^{-1}$ \citep{Schonrich2010} so this criterion should not
eliminate many M dwarfs (see Section \ref{sec.summary} for a
discussion of catalog completeness).

\begin{figure}
\includegraphics[width=84mm]{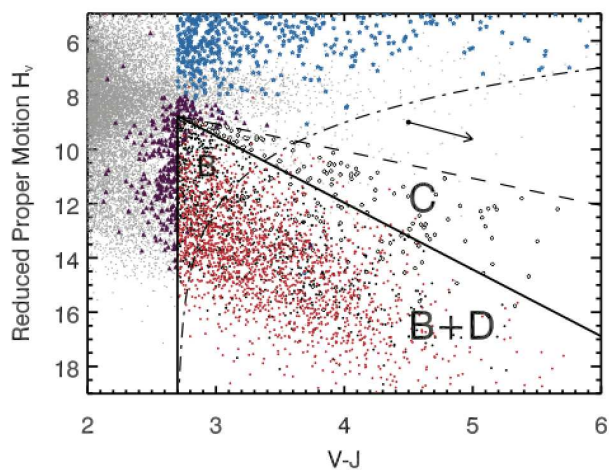}
\caption{2MASS $J$-$K_S$ color criterion for M dwarfs selected based on
  reduced proper motion.  Symbols are the same as in
  Figs. \ref{fig.absmag} and \ref{fig.redpm}.  The dotted line is the
  weighted mean value ($J$-$K_S=0.83$) for confirmed M dwarfs.  The
  solid lines mark $\pm$3 times the intrinsic dispersion in $J$-$K_S$
  remaining after formal errors are subtracted (0.028 magnitudes).}
 \label{fig.jk}
\end{figure}

To eliminate interloping giant stars, candidate M dwarfs were also
subjected to photometric color criteria developed using the colors of
bona fide M dwarfs identified by absolute magnitudes or spectroscopy.
We found that M dwarfs identified by absolute magnitude or spectrum
have a narrow range of $J$-$K_S$ colors compared to giant stars, with
a mean $J$-$K_S$=0.83, after eliminating outliers, and an intrinsic
dispersion of 0.028 magnitudes, after accounting for measurement error
(Fig. \ref{fig.jk}).  Stars with $J$-$K_S$ colors falling more than
three standard deviations from the locus (with photometry errors and
intrinsic dispersion added in quadrature) were excluded.

\begin{figure}
\includegraphics[width=84mm]{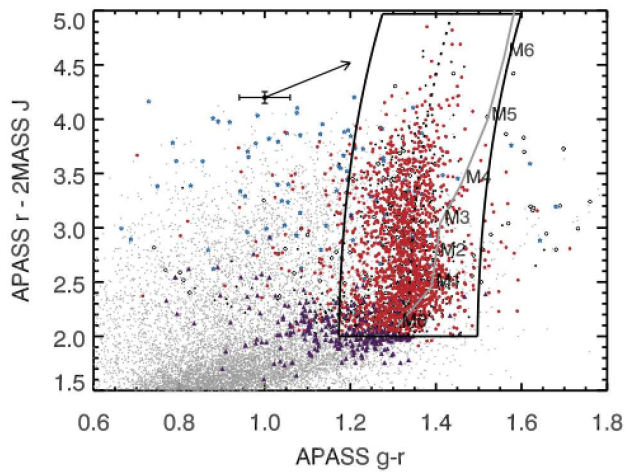}
\caption{APASS $g-r$ vs. $r-J$ color criterion for M dwarfs selected
  based on reduced proper motion.  Symbols are the same as in
  Figs. \ref{fig.absmag}, \ref{fig.redpm}, and \ref{fig.jk}.  The
  dotted line is a polynomial fit to values of confirmed M dwarfs and
  the solid lines denote the limits $r-J > 2$ and $\pm$3 times the
  intrinsic dispersion of the locus (0.054 magnitudes).  Mean SDSS
  colors of M dwarfs with different spectral types from
  \citet{Bochanski2011} are plotted as the grey line.}
 \label{fig.grj}
\end{figure}

We also applied criteria using $g-r$ and $r-J$ colors for those stars
where APASS photometry is available (Fig. \ref{fig.grj}).  We fit a
fifth-order polynomial to the median of $g-r$ values with low error
($< 0.03$ magnitudes) in $r-J$ bins.  The intrinsic dispersion about
this fit is 0.054 magnitudes.  Stars with $r-J < 2.7$ or $g-r$
magnitudes more than three standard deviations (photometric error and
intrinsic dispersion added in quadrature) were excluded.  The
trajectory of colors of M0-M6 dwarfs from the Sloan Digital Sky Survey
\citep[SDSS,][grey line in Fig. \ref{fig.grj}]{Bochanski2011} clearly
shows that APASS colors differ systematically from SDSS colors and
that the discrepancy increases with later spectral type.

We identified additional stars that had $M_V$ or $H_V$ values that are
brighter than the limits described above, but are still consistent
with the properties of M dwarfs if significant extinction or large
errors in $V$ or $J$ magnitudes are allowed, or they are especially
young and luminous, and/or are unresolved binaries.  USNO-B-based
errors in $V$ may be as large as one magnitude; error in $V$ is
represented by slope-one lines in plots of $M_V$ or $H_V$ vs. $V$-$J$
(Figs \ref{fig.absmag} and \ref{fig.redpm}).  There are six stars in
CONCH-SHELL with 2MASS $J$ magnitudes flagged as having low quality or
being upper limits: a few stars are bright enough that 2MASS
observations may have been in the detector's nonlinear regime or even
saturated.  That leads to more uncertain $J$ magnitudes which will
displace these stars along slope-one lines in Figs. \ref{fig.absmag}
and \ref{fig.redpm}.  In principle, interestellar or circumstellar
extinction could also displace stars off the main sequence locus.
Based on coefficients appropriate for the interstellar medium (ISM)
\citep{Yuan2013}, the direction of extinction/reddening (arrows) has a
slope of 1.25 in Figs. \ref{fig.absmag} and \ref{fig.redpm}.  Such
stars are also captured by this criterion.

On the other hand, distant and extincted giant stars or hotter dwarfs
would also be displaced to fainter magnitudes and redder colors along
the direction of the arrows in Figs. \ref{fig.absmag}-\ref{fig.grj}
and could contaminate the M dwarf catalog.  The $J<9$ magnitude limit
of the catalog plus the relationship between distance and interstellar
extinction limits this effect to stars with the bluest $V$-$J$ colors
in our sample.  We derived a relationship for the maximum plausible
reddening $E^*_{V-J}$ for a given observed $H_V$ and $V$-$J$, assuming
a linear relation between hydrogen column density $N_H$ and extinction
$A_V$.  We adopted $N_H = 2.21 \times 10^{21} {\rm cm}^{-2} A_V$
\citep{Guver2009}, characteristic of the diffuse ISM.  Thus for $A_V =
1.251 E_{V-J}$ \citep{Yuan2013}, the distance $D = 894\,{\rm
  pc}\,E_{V-J}\,n_H^{-1}$ where $n_H$ is the hydrogen space density of the
ISM in atoms~cm$^{-3}$.  In the $H_V$ vs. $V$-$J$ diagram the
magnitude limit $J < 9$ can be converted into an expression for
$E^*_{V-J}$ as a function of $n_H$, stellar transverse velocity $v_T$,
$H_V$ and $V$-$J$:
\begin{equation}
E^*_{V-J} = 0.26 \frac{n_H}{1~{\rm cm}^{-3}}\frac{v_T}{30~{\rm km~sec}^{-1}}10^{\frac{\left(V-J-2.7\right) - \left(H_v - 8.85\right)}{5}}
\end{equation}
The reference point $V-J = 2.7$, $H_V = 8.85$ is the blue, luminous
corner of the selection space in Fig. \ref{fig.redpm}, and 30
km~sec$^{-1}$ is the approximate stellar velocity dispersion at the
mid-plane of the galactic disk \citep{Bond2010}.  If the interstellar
medium along most lines of sight is characterized by $n_H \sim 1\,
{\rm cm}^{-3}$ then only stars with $V$-$J$ colors within about a
magnitude of 2.7 are potential contaminants and most stars are {\it a
  priori} unlikely to be interlopers.  We adopted a conservative
criterion of $5 \times E^*_{V-J}$ to define a ``danger zone'' in which
extincted contaminants might be found.  This is plotted as the
dot-dash line in Fig. \ref{fig.redpm} and bounds an envelope that
includes all stars identified as giants based on $M_V$.  M dwarf
candidates within the triangular region defined by this curve, the
$H_V$ limit, and $V-J > 2.7$ could be interlopers, especially if they
do not satisfy one or more of the criteria previously described.

We assigned M dwarfs or candidate M dwarfs to one of four classes
(A-D).  Stars in all four class have $J < 9$, $V-J > 2.7$, and
detectable proper motions.  The four classes, in order of decreasing
confidence, are:
\begin{itemize}
\item 1943 ``A-class'' stars which are spectroscopically confirmed M
  dwarfs in \citet{Lepine2013} or have absolute magnitudes that are
  not brighter than $3\sigma$ above the main sequence (solid and
  dotted curves in Fig. \ref{fig.absmag}) and at least 3$\sigma$
  fainter than $M_V = 4.2$, our criterion for evolved stars.  These
  are represented as the red points in
  Figs. \ref{fig.absmag}-\ref{fig.grj}.
\item 857 `B-class'' stars which do not have parallaxes but have
  reduced proper motions fainter than the selection limit represented
  as the solid lines in Fig. \ref{fig.redpm} and $J$-$K_S$, $g-r$, and
  $r$-$J$ colors, if available, within $3\sigma$ of the boundaries
  established for M dwarfs (solid lines in
  Figs. \ref{fig.jk}-\ref{fig.grj}).  These stars are represented by
  black points in Figs. \ref{fig.redpm}-\ref{fig.grj}.
\item 102 ``C-class'' stars which are not A- or B-class stars but have
  $M_V$ fainter than the dashed slope-one line in
  Fig. \ref{fig.absmag}, if parallaxes are available, or $H_V$ fainter
  than the dashed slope-one line in Fig. \ref{fig.redpm}, {\bf and}
  $J$-$K_S$, $g$-$r$, and $r$-$J$ colors, if available, that are
  consistent with M dwarfs.  These could be stars with large errors in
  $V$ or $J$ magnitudes, unresolved binaries, or young M dwarfs that
  are more luminous than the main sequence.  They are represented as
  open points in Figs. \ref{fig.absmag}-\ref{fig.grj}.
\item 93 ``D-class'' stars which are not A-, B-, or C-class stars,
  lack parallax measurements, have $H_V$ satisfying the dwarf
  selection criterion of B-class stars, but have $J$-$K_S$, $g-r$, or
  $r$-$J$ colors that are inconsistent with M dwarfs.  These could be
  dwarf stars that have errors in photometry, or are flaring or
  rotationally variable stars where the photometry in different
  bandpasses was obtained at different epochs.  To avoid reddened
  hotter or giant stars, objects inside the ``danger zone'' of
  Fig. \ref{fig.redpm} were not included.  D-class stars are also
  represented by open points in Figs. \ref{fig.redpm}-\ref{fig.grj}.

\end{itemize}

The total number of confirmed or candidate M dwarfs in our initial
catalog is 2995.  Of these 532 are not in LG11, and there are 319 LG11
stars with $J<9$ which were not selected, mostly because the revised
$V$ magnitudes (e.g., APASS replacing USNO-B) are brighter and thus
$V-J$ becomes bluer than 2.7.  We describe a comparison with the F13
catalog in Section \ref{sec.frith}.  Based on spectra of these stars
we eliminated 44 stars (Section \ref{sec.observations}), leaving a
final catalog of 3007 stars.

\section{Spectroscopy}
\label{sec.observations}
\subsection{Observations}

Visible-wavelength spectra with a resolution $R = \lambda/\Delta
\lambda \sim 10^3$ were obtained with the SuperNova Integral Field
Spectrograph (SNIFS) on the University of Hawaii 2.2~m telescope on
Maunakea, Hawaii, the Mark III spectrograph and Boller \& Chivens CCDS
spectrograph (CCDS) on the 1.3~m McGraw-Hill telescope at the MDM
Observatory on Kitt Peak, Arizona, the REOSC spectrograph on the
2.15~m Jorge Sahade telescope at the Complejo Astron\'{o}mico El
Leoncito Observatory (CASLEO), Argentina, and the RC spectrograph on
the 1.9~m Radcliffe telescope at the South African Astronomical
Observatory.  SNIFS acquires 3200-9700\AA{} integral field spectra in
blue and red channels that narrowly overlap at 5100-5200\AA{}
\citep{Lantz2004}.  Except for metallicity determination (Section
\ref{sec.metallicity}), only the red channel data were used in this
project as many stars had very low signal-to-noise in the blue
channel.  The Mark III was used with a 1.52 arcsec slit, a Hoya yellow
order-separation filter, either a 300 or 600 lines mm$^{-1}$ grating
blazed at 5800\AA{} and either the ``Wilbur'' or ``Nellie'' 2048$^2$
CCD detectors.  The CCDS was used with a 158 lines mm$^{-1}$ grating
blazed at 7530\AA{} and a 1 arcsec slit, and spectra from this
instrument cover 4800-8800\AA{}.  The REOSC spectrograph was used with
a 300 lines~mm$^{-1}$ grating blazed at 5000\AA{} and is equipped with
a $1024^2$ TEK CCD which is thinned and back-illuminated.  The RC
Radcliffe spectrograph was used with a grating having 300 lines
mm$^{-1}$ blazed at 7800\AA{} for a dispersion of 3.15\AA{}
pixel$^{-1}$ on a SITe 1024$^2$ CCD.

We obtained a total of 3071 spectra of 2583 stars or 86\% of the
catalog over the span of more than 11 years. 425 stars were observed
twice, 14 stars were observed thrice, and 6 stars had more than four
observations.  A summary of the observations with each
telescope/instrument combination is presented in Table
\ref{tab.telescopes}.

\subsection{Reduction}

{\it UH~2.2m and SNIFS:} The majority of SNIFS data reduction was
performed with the SNIFS data reduction pipeline, which is described
in detail in \citet{Bacon2001} and \citet{Aldering2002}.  To
summarize, the SNIFS pipeline performed standard CCD processing (i.e.,
dark, bias, and flat field corrections), and then assembled the data
into two data cubes for the red and blue channels. Each data cube was
then cleaned of cosmic rays and bad pixels.  To mitigate errors from
telescope flexure, the data were wavelength-calibrated using arc lamp
exposures acquired immediately after the science exposure.  The SNIFS
pipeline then used a point-spread function model to estimate and
subtract the background, and extract the 1D spectra from the data
cube.  For $\la 1$\% of sources the extraction failed, usually due to
the presence of a marginally resolved binary, unusually high seeing
($\ge3$\arcsec), or a software failure. In these cases we identified
the star position and extracted the 1D spectrum manually.  An
approximate flux calibration was applied by the SNIFS pipeline to each
spectrum using an approximate instrument and atmospheric response
function.

We applied an additional correction to the flux calibration using our
own model.  During each night we observed two to five standards with
well-calibrated spectra from \citet{Oke1990}, \citet{Hamuy1994},
\citet{Bohlin1995}, \cite{Bessell1999}, or \citet{Bohlin2001}.  We
derived an empirical wavelength- and airmass-dependent correction by
comparing the spectra of all standard star observations taken over the
course of the project to their spectra in the literature.  We further
derived a nightly term by the same technique using just the standard
stars observed in a given night.  However we found that the
night-dependent correction was not significant on photometric nights.
This method enabled us to avoid the impractical task of obtaining
spectra of standards spanning the full range of observed airmasses
each night. \citet{Mann2013c} found that synthetic photometry from
SNIFS spectra is in excellent agreement with colors from ground-based
photometry, suggesting systematic errors in the flux calibration are
small.  As an additional test, we compared observations of the same
star on different nights.  Our results suggest that random errors in
the flux calibration are $\la1$\%, except around the atmospheric
\water{} band at 9300-9600\AA, which is known to vary on timescales
shorter than those considered by our model. UH2.2m and SNIFS have been
shown to be stable at the $<0.1\%$ level over the course of hours
\citep{Mann2012b}, which suggests that the additional noise is coming
from errors in the extraction process \citep[see ][for a more detailed
  discussion]{Buton2013}.

{\it MDM and Mark~III or CCDS:} Reduction of most MDM spectra were
performed using the IRAF reduction package\footnote{IRAF is
  distributed by the National Optical Astronomy Observatory, which is
  operated by the Association of Universities for Research in
  Astronomy (AURA) under cooperative agreement with the National
  Science Foundation.}  Images were de-biased and flat-fielded using
the CCDPROC package. Sky emission was then subtracted, and star
spectra extracted using the DOSLIT routine in the SPECRED
package. Wavelength calibration was performed using arc line spectra
of Ne+Ar+Xe lamps which were routinely collected after each visit on a
target, to account for flexure in the spectrographs. In a small number
of cases in which arc spectra were not collected immmediately after
the visit, calibration was performed using the arc lamp collected on
the following target, thus potentially producing small but systematic
shifts.  Flux calibration was performed with observations of the
calibration standard stars Feige 110, Feige 66, Feige 34, and Wolf
1346 \citep{Oke1990}, either one of which was typically observed once
every night during oberving runs.  Spectra collected with the CCDS
spectrograph were imaged with thin CCDs and displaying significant
fringing redward of 7000\AA{}. In this case, additional flatfields
were collected immediately after each visit on a star, either just
before or just after calibration arc lamps were collected. In a few
instances, however, additional flatfield lamps were not collected due
to overlook on the part of the observer. Flat lamps from similar
H.A./Dec. pointings were used to correct for fringing, but this
sometimes failed to completely eliminate the fringing patterns. As a
result, weak fringing features are often seen redward of 7500\AA{} in
some spectra.  MDM spectra were also occasionally found to be affected
by slit losses due to our use of a relatively narrow slit and to
observations conducted at large hour angles ($<$3 hours from meridian)
with the slit oriented north-south of the sky and not strictly
oriented along the local parallactic angle. In some cases, it was
possible to determine the pattern of the slit loss based on
observations of the calibration stars, and flux recalibrations were
performed to correct for the losses.

We obtained spectra of several bright dwarfs to calibrate our
estimates of effective temperature (Section \ref{sec.params}).  Some
of these stars are in the CONCH-SHELL itself.  These stars were
reduced and calibrated using IDL scripts, rather than IRAF.  Spectral
images were debiased, flattened by quartz lamp flats, and the source
spectrum traced by a fitting a third-order polynomial to centroid
positions vs. wavelength.  Cosmic ray events were identified by their
effect on the along-slit width of the spectral image and filtered.
Sky spectra were extracted from two flanking apertures and subtracted
from the raw source spectrum.  Quadratic pixel-wavelength solutions
were derived from the arc lamp spectrum (Ne, Ar, and/or Xe) acquired
closest in time to the target.  CCDS spectra exhibit severe fringing
at the red end, which required that flat fields be obtained at each
pointing.  This step did not entirely remove the fringes and it was
necessary to identify the fringe pattern in each spectrum by Fourier
transform and smoothing at the peak spatial frequency to remove the
pattern.  Extinction correction and flux calibration were performed
using the standard KPNO extinction table and the spectrophotometric
standards Feige 34, 66, and 110 \citep{Oke1990}.

{\it CASLEO and REOSC:} Long-slit spectra were obtained with the REOSC
spectrograph by replacing the echelle grid by a mirror
\citep[see][]{Cincunegui2004}.  For each star we obtained two spectra
to help us to remove cosmic rays.  Each of the two spectra were bias
corrected, optimally extracted and wavelength-calibrated using
standard IRAF routines.  The wavelength calibration was performed
using Cu-Ar arc lamp spectra.  Then we combined both spectra, removing
cosmic rays.  To calibrate these spectra in flux, we also observed
each night at least four standard stars selected from the Catalogue of
Southern Spectrophotometric Standards \citep{Hamuy1994}. The reduction
and calibration were performed using standard IRAF routines.

{\it SAAO Radcliffe and RC:} Reduction of the spectra was performed
closely following the prescription detailed in \citet{James2013}, with
the exception that all processing was executed within the IRAF
environment \citep{Tody1993} instead of the Starlink\footnote{Please
  see http://starlink.jach.hawaii.edu} one. Extracted and
wavelength-corrected spectra for all targets and calibrator standard
stars were corrected for local atmospheric extinction using an updated
version of the \citet{SpencerJones1980} study. On a per night basis,
extinction-corrected count rates were converted to flux by reference
to the spectrophotometric flux standard star Feige 110, and its
tabulated values in \citet{Massey1988} and \citet{Massey1990}.  Flux
calibration for spectra acquired on the night of UT 2013 Sept. 22 was
performed using the spectrum of Feige~110 obtained on the night of
Sept. 21 due to a case of mistaken identity of the flux standard
observed on Sept 22.

{\it Flux normalization:} Many spectra were not obtained under
conditions where absolute flux normalization was possible.  Instead,
all spectra were normalized to SDSS $i=0$.  This band was chosen
because it is mostly or entirely covered by our spectra and the
emission from early M dwarfs peaks at $i$-band.  For SNIFS spectra the
O$_2$ telluric lines at $\sim7600$\AA\ was removed as part of our
reduction, but this was not the case for most spectra obtained with
other spectrographs. We derive an approximate correction to the
telluric features in the remaining spectra using observations of white
dwarfs or hot stars taken with the appropriate spectrograph. We
assumed these stars were smooth (no features) around the O$_2$
atmospheric lines and therefore we measured the strength of the
telluric lines by comparing the observed spectrum to one interpolated
from the uncontaminated parts of the spectrum. Because telluric line
strengths vary as a function of atmospheric conditions and observed
airmass this correction is only approximate. After we applied the
telluric correction, we convolved each spectrum with the SDSS $i$
filter transmission
profile\footnote{http://www.sdss.org/dr7/instruments/imager/filters/i.dat}. We
then integrated over the resulting spectrum, and calculated a
synthetic magnitude using the zero-points from
\citet{Fukugita1996}. The synthetic magnitude was used to calculate a
normalization constant such that the spectrum becomes that of a star
with $i = 0$.

\section[]{Stellar Properties}
\label{sec.parameters}

\subsection{Luminosity Class and Spectral Type}

\begin{figure}
\includegraphics[width=84mm]{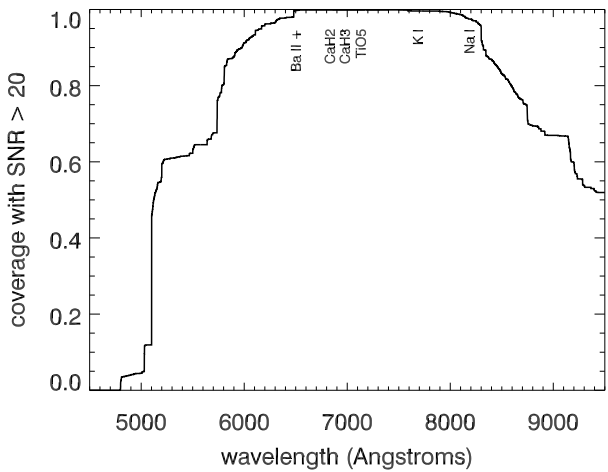}
\caption{Wavelength coverage of our spectra at signal-to-noise per
  resolution element $>20$.  The locations of four gravity-sensitive
  indices calculated to identify interloping giant stars plus TiO5, a
  proxy for effective temperature, are shown.}
 \label{fig.coverage}
\end{figure}

We calculated four gravity-sensitive indices at wavelengths between
6400 and 8300\AA, a region covered by nearby all of our spectra
(Fig. \ref{fig.coverage}).  These indices were (i) the averaged CaH2
and CaH3 indices (ratio of flux in bands at 6830\AA{} and 6975\AA{} to
the continuum), (ii) the equivalent width of the K~I line at 7699\AA,
(iii) the equivalent width of the Na~I line at 8185\AA, and the
equivalent width of a blend of Ba~II, Fe~I, Mn~I, and Ti~I lines
centered at 6500\AA{}.  \citet{Mann2012} found these indices to be
effective discriminators between M dwarfs and giants with
moderate-resolution spectra.  We also calculated the TiO5 index,
centered at 7130\AA{}, an indicator of effective temperature for M
dwarfs \citep{Reid2005}.  The band and continuum definitions in
\citet{Mann2012} were used, with the exception of Na~I (see below).
Each spectrum was shifted by the offset found between its vacuum
wavelength version and the rest-frame predicted spectra of a best-fit
PHOENIX atmosphere model (see Section \ref{sec.params}).  Errors for
each index were calculated by Monte Carlo simulations that included
both the formal noise in each spectrum plus an assumed error in
wavelength calibration with an RMS of 0.5\AA.  Fifty-two CASLEO/REOSC
spectra were obtained with an incorrect grating setting and lack the
region around \ha{} and the Ba II feature.

\begin{figure}
\includegraphics[width=84mm]{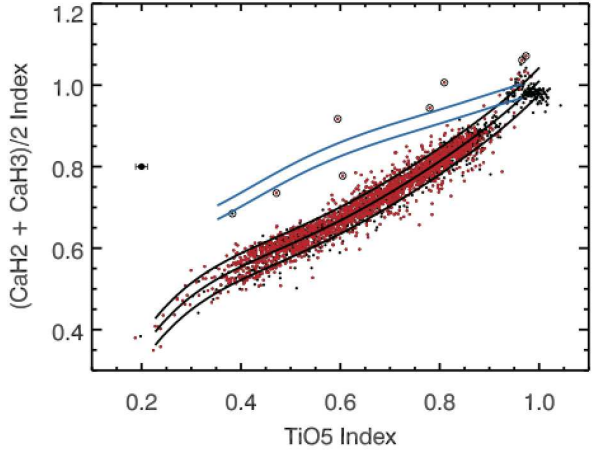}
\caption{Mean of the CaH2 and CaH3 indices vs. TiO5 index, an
  indicator of effective temperature in M dwarfs.  The isolated point
  on the left illustrates the median errors.  Red points are M dwarfs
  confirmed by parallaxes or previous spectroscopy.  The solid black
  lines are a least-squares polynomial fit to the M dwarf locus and
  plus and minus twice the intrinsic standard deviation around the
  locus, after subtraction of formal measurements error.  Circled
  points are spectra where the CaH index is $>5\sigma$ above the
  locus, where $\sigma$ is the measurement error and intrinsic locus
  width added in quadrature.  The blue lines are the $\pm 2\sigma$
  contours for the sample of giant stars constructed by
  \citet{Mann2012}.}
 \label{fig.cah}
\end{figure}

\begin{figure}
\includegraphics[width=84mm]{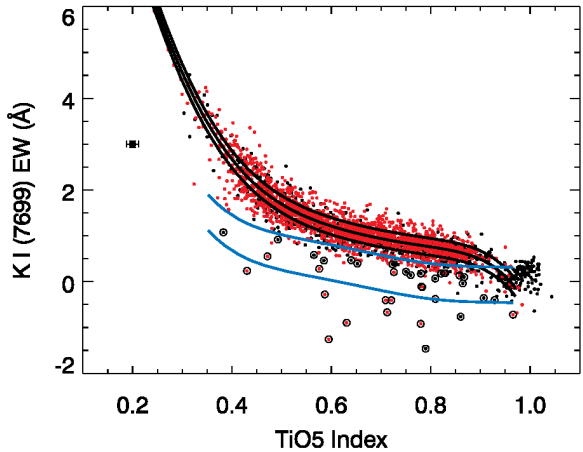}
\caption{Same as Fig. \ref{fig.cah} except for the equivalent width of
  the K~I line at 7699\AA.}
 \label{fig.ki}
\end{figure}

\begin{figure}
\includegraphics[width=84mm]{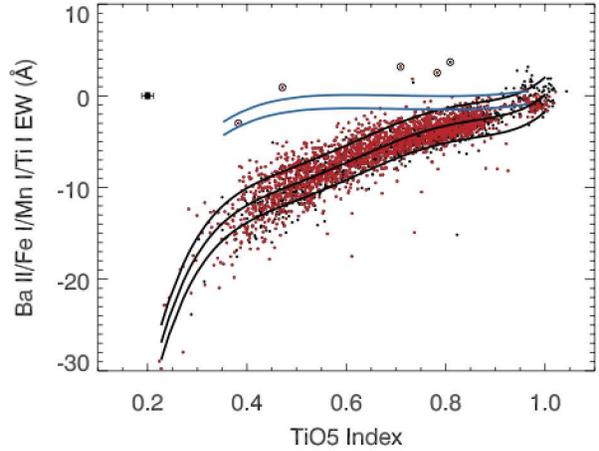}
\caption{Same as Fig. \ref{fig.cah} except for the equivalent width of
  a blend of lines of Ba~II, Fe I, Mn~I, and Ti~I at $\sim$6500\AA..}
 \label{fig.baii}
\end{figure}

\begin{figure}
\includegraphics[width=84mm]{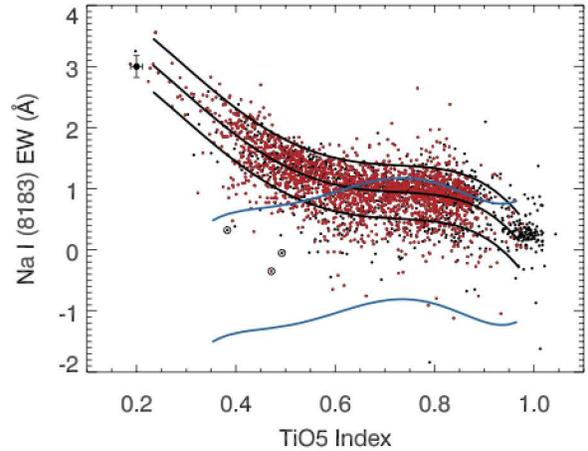}
\caption{Same as Fig. \ref{fig.cah} except for the equivalent width of
  the Na~I line at 8183\AA.}
 \label{fig.nai}
\end{figure}

The four gravity-sensitive indices are plotted vs. the TiO5 index in
Figs. \ref{fig.cah}-\ref{fig.baii}.  Stars with TiO5 index $< 1$ are
M-type while those with TiO5 $\approx 1$ are mostly late K stars but
could include earlier spectral types as well.  For each index we fit a
polynomial with TiO5 to the locus and calculated the intrinsic scatter
around the locus after subtracting the measurement errors.  We found
that the EW of the Na~I doublet as defined in \citet{Schiavon1997} and
used by \citet{Mann2012} produces a very large scatter, probably
because the line at 8195\AA{} and the continuum region redward of this
is beyond the useful wavelength range of many of our spectra or,
possibly, the presence of uncorrected telluric lines.  Instead, we
measured the EW of the 8183\AA{} line in the range 8172-8197\AA{} and
only used the blue continuum region (8170-8173\AA) defined in
\citet{Schiavon1997}.  This reduced the scatter in EW, although it is
still larger than that of the other lines (Fig. \ref{fig.nai}).  The
Na~I line is especially sensitive to metallicity \citep{Mann2013a} and
this may partly explain the larger scatter.

We flagged 39 spectra with K~I, Ba~II+, or CaH indices at least
$5\sigma$ below the best-fit locus (circled points in
Figs. \ref{fig.ki}-\ref{fig.nai}).  However, 20 of these are of
A-class stars confirmed by parallaxes and/or spectra in
\citet{Lepine2013} (red points), including some observations of very
bright calibrator M dwarfs which may have entered the nonlinear
response regime of the MDM Mark III detector.  The majority of the
flagged stars do not fall within the giant locus bounded by blue lines
in Figs. \ref{fig.ki}-\ref{fig.nai}, also suggesting a problem with
the spectra rather than the that these are giants.  The two most
likely interlopers among the 39 flagged stars is the C-class star
PM~I13193-5800 (Tycho 8657-739-1), and one D-class star PM~I06298-2250
(Tycho 6507-473-1).  A SIMBAD search revealed no specific published
information on any of these stars.

Fourteen of the 20 flagged spectra affiliated with A-class stars and
18 of the 19 flagged spectra belonging to B-, C- or D- class stars are
flagged exclusively because of weak K~I lines.  Twenty-four of these
were obtained with the REOSC at CASLEO and may be the product of
wavelength calibration error, truncation of the spectra due to an
incorrect grating setting, contamination by brighter nearby stars, or
clouds.  Only 10 of the flagged spectra produce a weak Ba II or CaH
index, and 9 of these are A-class stars, all established M dwarfs.
Some of these spectra are clearly saturated or are contaminated by
much brighter, solar-type companions.  One curious case is the
high-proper motion M4 dwarf GJ~1218, for which all four indices are
weak.  This star may be metal-poor although its luminosity ($M_V =
11.71$) rules out a subdwarf classification.  The last spectrum is
that of the D-class star PM~I06298-2250, also with four weak indices,
and it is probably of a giant: we excluded this star from the catalog.

\begin{figure}
\includegraphics[width=84mm]{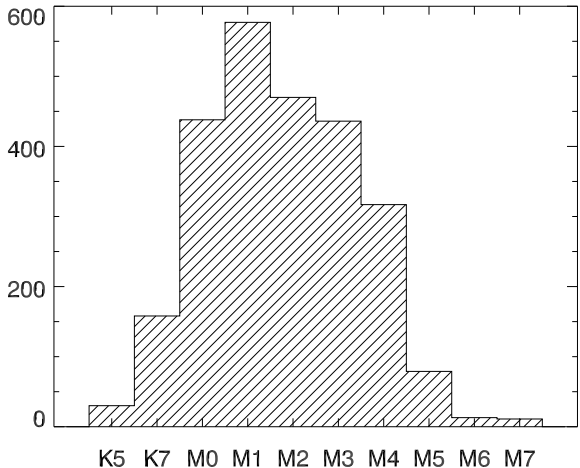}
\caption{Distribution of CONCH-SHELL spectral types as determined by
  HAMMER in manual mode.  An additional 18 stars in the catalog have
  spectra but uncertain spectral types and are not included.}
 \label{fig.sptype}
\end{figure}

Spectral types were determined using HAMMER \citep{Covey2007}.
Because of a systematic error in the HAMMER's automated spectral
typing \citep{Lepine2013} we used manual assignments.  The
distribution of spectral types is plotted in Fig. \ref{fig.sptype}.
We were unable to assign spectral types to 2 stars: Another 59 stars
have spectra that appear to be earlier than K5.  However, we could not
manually assign accurate spectral types with HAMMER due to the lack of
obvious spectral features, the minimal overlap between the HAMMER
templates and these spectra, uncorrected slit losses, and/or other
problems with the spectra.  Among these 61 stars are 8 A-class stars,
all of which are established M dwarfs or proper-motion stars according
to SIMBAD.  All of the TiO5 indices affiliated with these spectra are
$>0.92$, consistent with stars earlier than M, but this does not
exclude late K stars.  The distributions of these objects with $V$-$J$
color and galactic latitude $b$ include a cluster at $V$-$J<3.4$ and
$|b| < 6$~deg.  We removed all B-class stars in this cluster, as well
as all C- and D-class stars among the 61 (5 stars in total).  Thus we
excluded a total of six stars based on their spectra, leaving 2989
stars.  The overall contamination rate by giants and hotter stars
before spectroscopic screening is $<$1\%.  

\subsection[]{Metallicity}
\label{sec.metallicity}

Metallicities with respect to the solar value ([Fe/H]) were estimated
following the method of \citet{Mann2013a}.  They used FGK+M wide
binaries to identify metal-sensitive atomic and molecular features in
M dwarf spectra, from which they derived an empirical calibration
between the strength of these features and the metallicities of the
late-type dwarf. We calculated metallicities only for CONCH-SHELL
stars with SNIFS spectra. This was done because the \citet{Mann2013a}
calibration utilizes at least one feature blueward of 4800\AA, which
is only covered by our SNIFS spectra, and because the calibration of
\citet{Mann2013a} was itself derived from SNIFS spectra.

A total of 1338 stars were observed using SNIFS. We removed 100 stars
because they have spectral types outside the range where the
calibration is valid (K7-M5) and 53 stars because their SNR in the
blue channel is too low ($<30$).  We placed each of the remaining 1185
spectra in their rest frames by converting wavelengths to vacuum
values then cross-correlating each spectrum to SDSS templates
\citep{Bochanski2007} of the corresponding spectral subtype. We then
used an IDL routine\footnote{https://github.com/awmann/metal} to
calculate the metallicity of each star. Errors in [Fe/H] are
calculated by combining (in quadrature) measurement errors and
calibration errors reported by \citet{Mann2013a}.

\begin{figure}
\includegraphics[width=84mm]{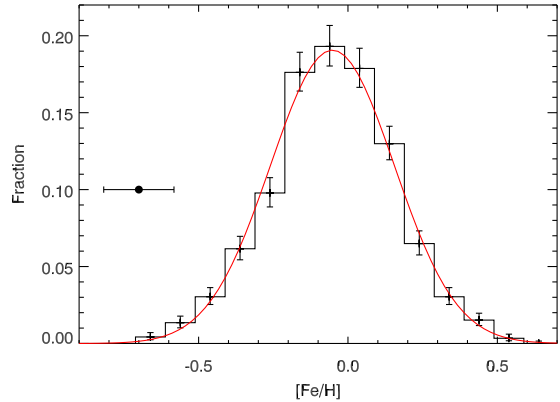}
\caption{Distribution of metallicity among 1185 stars with spectra
  obtained by SNIFS.  Poisson errors in each bin are shown. The lone
  point denotes the median error in [Fe/H] for an individual star. A
  best-fit gaussian is shown in red. The resulting fit is centered at
  [Fe/H] = -0.07~dex with a width ($\sigma$) of 0.20~dex. }
\label{fig.metalhist}
\end{figure}

The resulting distribution of metallicities is plotted in
Fig.~\ref{fig.metalhist}. The distribution is well described by a
Gaussian centered at [Fe/H] = -0.05 with a standard deviation of
0.21~dex.  The intrinsic width, after correction for measurement
error, is 0.18 dex. This is consistent with previous estimates of
volume-limited M dwarf samples \citep{Johnson2009,Schlaufman2010}, and
very similar to the distribution of FGK stars in the solar
neighborhood \citep[median metallicity = -0.06, standard deviation =
  0.21~dex,][]{Casagrande2011}.

\subsection{Physical Parameters}
\label{sec.params}

To estimate the effective temperature \teff{}, radius $R_*$,
luminosity $L_*$, and masses $M_*$ of these M dwarfs we followed the
procedure of \citet{Mann2013c}, first determining \teff{} by finding
the best-fit model stellar spectrum, then using the best-fit
temperature in empirical relations to determine the other parameters.
This procedure was calibrated on nearby stars with measured radii,
distances and bolometric fluxes, and hence bolometrically-determined
temperatures \citep{Boyajian2012}.  Flux-calibrated,
extinction-corrected spectra were compared with the predictions of the
BT-SETTL version of the PHOENIX stellar atmosphere model
\citep{Rajpurohit2014}.  We employed the suite of models with
\citet{Caffau2010} solar abundances.  \citet{Mann2013c} showed that
minimum-$\chi^2$ fitting of a grid of models and their interpolations
recovered the bolometric temperatures of M dwarfs with an accuracy of
60~K.

We followed the procedure of \citet{Mann2013c}, with a few
modifications.  We excluded the same set of wavelength intervals where
the models perform poorly to improve the fit.  The observed and model
spectra are normalized by their median values, and a uniform
wavelength offset between them is allowed as a free parameter of the
fit.  However, we introduced a third-order polynomial with wavelength
to represent slit loss: the coefficients are free parameters and are
not interpreted.  We also added a quadratic term with model [Fe/H] to
the $\chi^2$ used to describe the goodness-of-fit of a model,
i.e. $({\rm [Fe/H]-[Fe/H]}_0)^2/\sigma_{\rm [Fe/H]}^2$.  For fits to
SNIFS spectra where the stellar metallicity was determined (Section
\ref{sec.metallicity}), [Fe/H]$_0$ is the measured value and
$\sigma_{\rm [Fe/H]}^2$ is the measurement uncertainty.  For fits to
other spectra, we substituted the mean and standard deviation of all
SNIFS values of [Fe/H].  To more thoroughly explore the range of
possible spectra, 10000 interpolations were generated from sets of
{\it three} rather than two normalized spectra.  The interpolations
draw from the best-fit (minimum $\chi^2$) model spectrum and at least
6 other model spectra with the lowest $\chi^2$, up to
$\chi^2_{min}\left(1 + \Delta \chi_{\nu}^2\right)$, where $\Delta
\chi_{\nu}^2$ is the increase in the reduced $\chi^2$ corresponding to
the 95\% confidence interval.  After the best fit of these
interpolations was identified, we estimated the error in \teff{}
calculated as one-fourth the 95\% confidence interval in $\chi^2$.  We
added 60K in quadrature to this error to represent the accuracy of our
calibration.

\begin{figure}
\includegraphics[width=84mm]{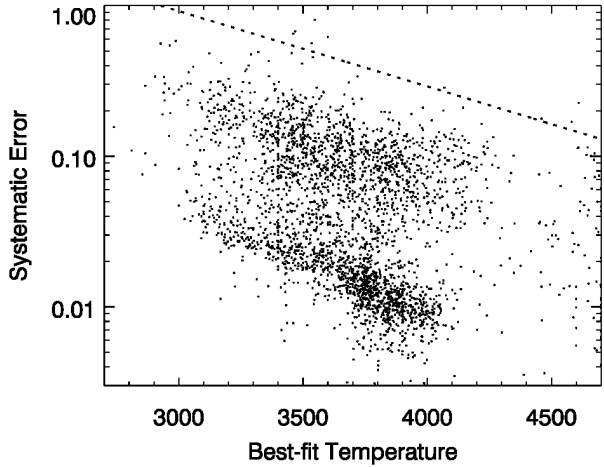}
\caption{Systematic error $\epsilon$ of the best-fit of PHOENIX
  BT-SETTL model to stellar spectra vs best-fit \teff{}.  Fits with
  $\epsilon$ above the dashed line were not used to calculate stellar
  parameters.}
 \label{fig.fiterr}
\end{figure}

For each best-fit model we calculated the associated quantity
$\epsilon = \sqrt{\chi_{\nu}^2 -1}/{\rm SNR}$, an estimator of the
mean error in excess of formal error due to observational systematics
and imperfect modeling of the stellar spectrum.  Values of $\epsilon$
are plotted vs. the best-fit \teff{} in Fig. \ref{fig.fiterr}.  The
locus of points at low $\epsilon$ are dominated by SNIFS spectra, with
systematic errors as low as $\sim$1\%.  This locus rises with
decreasing \teff{} as the imperfectly-modeled molecular bands of M
dwarf spectra become more pronounced and contribute more to
$\chi^2_{\nu}$.

Forty-one spectra with $\epsilon$ above the dashed line in
Fig. \ref{fig.fiterr} and/or best-fit wavelength offsets exceeding the
spectrograph FWHM, indicating a problem with fitting or wavelength
calibration, were not used to estimate \teff{}.  An offset exceeding
5.4\AA{}, the highest spectral resolution in our survey, corresponds
to a radial velocity of 250~km~sec$^{-1}$, something exceedingly
unlikely to be observed in our sample.  Some CASLEO spectra were
obtained with an improper grating setting and that limited the range
of usable wavelengths.  Others were obtained at high airmass or on
cloudy nights, with low signal-to-noise, or suffered contamination by
twilight or a nearby full Moon.  If more than one acceptable value of
\teff{} was available a weighted mean was used.

\begin{figure}
\includegraphics[width=84mm]{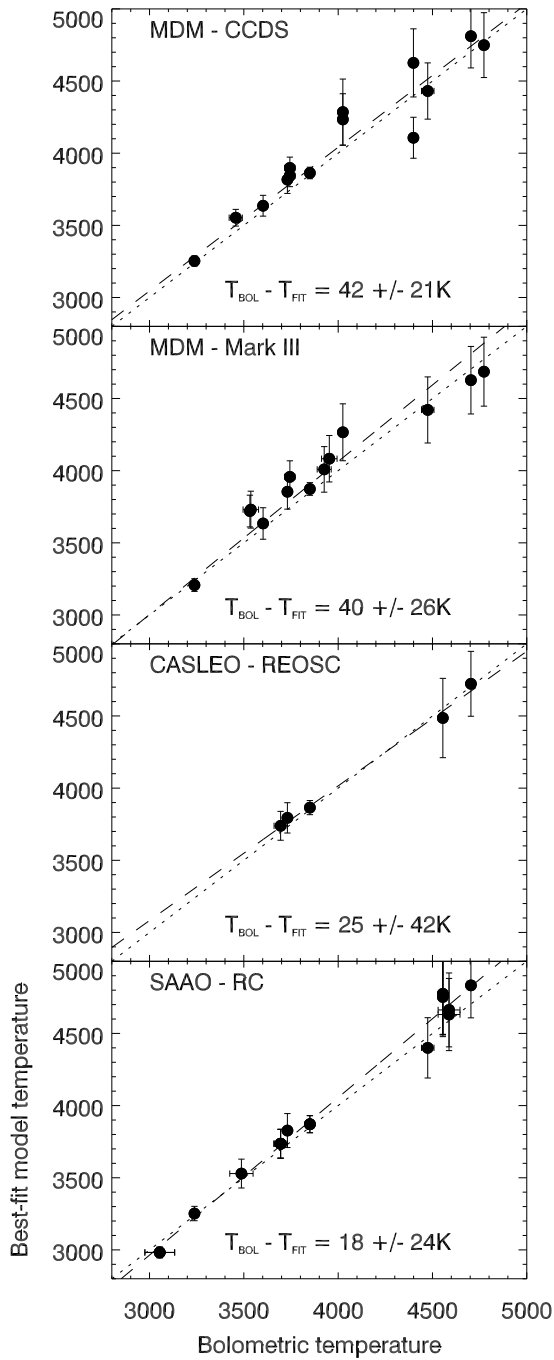}
\caption{Stellar \teff{} of calibrator stars from least-squares
  fitting of PHOENIX models to spectra obtained with four instruments
  on three telescopes.  These values are compared to bolometric
  temperatures determined from measurements of angular radii and
  bolometric fluxes.  The dotted line is equality and the dashed line
  is a minimum $\chi^2$ fit.  The weighted-mean offset is reported in
  each panel.}
 \label{fig.tempcal}
\end{figure}

\citet{Mann2013c} calibrated this method of obtaining \teff{} using
spectra obtained with SNIFS.  The same analysis of spectra obtained
with other instruments may introduce systematic differences between
the best-fit temperatures and the bolometric temperatures.  To
determine such offsets we observed several of the \citet{Boyajian2012}
calibrator stars having \teff{}$<4200$~K with each
telescope/instrument combination.  The comparisons between best-fit
and bolometric temperatures are shown in Fig. \ref{fig.tempcal}.  We
calculated the weighted mean offset for each instrument: an F-test
using the ratio of variances showed that fitting a line with a
non-unit slope did not significantly improve the fit.  The offsets
($T_{\rm bol}-T_{\rm fit}$) were $40 \pm 26$K for the Mark III at MDM,
$42 \pm 21$K for the CCDS at MDM, $25 \pm 42$ for the CASLEO/REOSC
spectrograph, and $18 \pm 24$ for the SAAO Radcliffe~1.9m/RC
spectrograph.  Spectra for most of the stars in the northern
hemisphere were previously analyzed and temperatures estimated by the
same process of model fitting \citep{Lepine2013} but using BT-SETTL
models with the \citet{Asplund2009} abundances.  Our revised
temperatures using the new PHOENIX models are systematically
100-150K hotter, as was previously noted by \citet{Mann2013c}.

Models of the spectra of M dwarf stars, particularly the TiO and CaH
lines, have significantly advanced, but challenges remain
\citep{Rajpurohit2014}.  Discrepancies between the models and the
actual spectra of stars will (i) inflate contribution of the spectral
contributions to $\chi^2$ relative to other measurements such as
[Fe/H] and (ii) bias stellar parameters from least-$\chi^2$ fits
toward the direction of more reliable stellar models, not necessarily
more accurate parameters.  The trend of increasing systematic error
with decreasing \teff{} in Fig. \ref{fig.fiterr} raises the spectre of
a bias in best-fit \teff{} towards higher temperatures where the
PHOENIX models are more accurate.  However, the excellent agreement
between best spectral fit temperatures and bolometric temperatures of
our calibrator stars (Fig. \ref{fig.tempcal}) indicates this effect is
small, perhaps limited by the deep features in the spectra of late M
dwarfs.

\begin{figure}
\includegraphics[width=84mm]{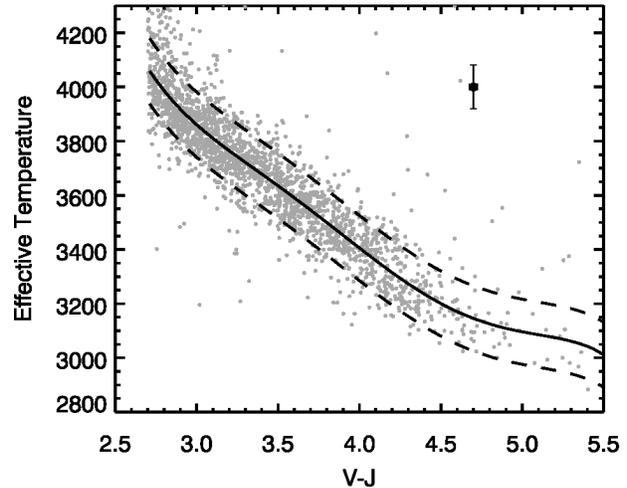}
\caption{Stellar \teff{} of stars from least-squares fitting of
  PHOENIX models vs. $V$-$J$ color.  The single point with the error
  bars represents median uncertainties in $V$-$J$ and \teff{} (91K).
  The curve is the weighted least-squares fit of a third-order
  polynomial and is used to estimate the \teff{} of stars without
  spectra or acceptable PHOENIX model fits.  The dashed lines
  represent $\pm$ twice the intrinsic width of the locus (40K) in
  \teff{} after accounting for formal errors.}
 \label{fig.teff-color}
\end{figure}

If no value from acceptable spectral fits was available, we calculated
\teff{} based on $V$-$J$ color and a best-fit polynomial for \teff{}
vs. $V$-$J$ (Fig. \ref{fig.teff-color}): $T_{\rm eff} = 4068.8 -
916.1\left(V-J-2.7\right) + 956.8\left(V-J-2.7\right)^2 -
868.0\left(V-J-2.7\right)^3$.  The uncertainties in these values are
derived by adding in quadrature the uncertainties from error in
$V$-$J$, the intrinsic scatter of the locus (40~K), and the
uncertainty in the zero point of the spectroscopic \teff{} calibration
\citep[43~K,][]{Mann2013c}.  A number of stars yield spectral best-fit
\teff{} values that are significantly hotter than their $V-J$ colors;
these conflicts may be the result of blends with brighter stars
(affecting photometry) but also inaccuracies of spectral fits at
\teff{}$>4200$K where there are no broad molecular features.  For
cases of the latter kind we replaced the spectroscopic \teff{} with
values based on $V-J$.  Nineteen D-class stars with $V-J > 6$, beyond
the valid range of our fit --- and the plausible range of bright M
dwarfs --- were excluded.  We retained one M dwarf with $V-J > 6$
(GJ~1230B or PM~I18411+2447N), but did not assign a value of \teff{}.
Our final catalog contains 2970 stars.

\begin{figure}
\includegraphics[width=84mm]{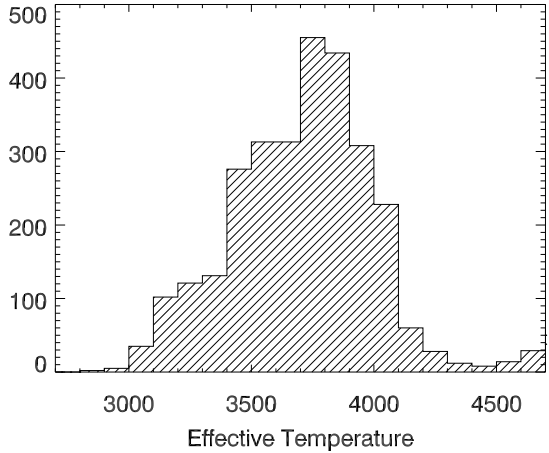}
\caption{Distribution of stellar \teff{} in the CONCH-SHELL catalog.
  Values for stars without acceptable spectra are based on $V$-$J$
  color.}
 \label{fig.teff}
\end{figure}

The distribution of the estimated \teff{} values of CONCH-SHELL stars
is plotted in Fig. \ref{fig.teff} and included in Table
\ref{tab.catalog}.  The nearly total absence of stars cooler than
\teff{} $\sim 3000$K is a result of the magnitude limit of the
catalog.  The appearance of stars hotter than \teff{} $\sim 4000$K
reflects the dispersion between stellar colors and \teff{} and the
inclusion of late K dwarfs in this catalog, plus errors exceeding
100~K for many stars.

We estimated stellar radius $R_*$, luminosity $L_*$, and mass $M_*$
using the metallicity-independent empirical relations of
\citet{Mann2013c}.  Our calibration is only valid for \teff{}$>$3238K.
For cooler stars we report upper limits based on \teff{} = 3238K.
These values are reported in Table \ref{tab.catalog}.  Errors were
calculated by combining, in quadrature, the formal errors from the
uncertainty in \teff{} and the uncertainties in the empirical
calibration.

\subsection{Comparison of Parallax- and Spectroscopy-Based Luminosities and Masses}

For some stars we have parallaxes which, along with a bolometric
correction, allowed us to independently determine luminosities and
estimate masses from a mass-luminosity relation.  We constructed a
bolometric correction to the $J$-band magnitudes of M dwarfs using the
parameters in \citet{Mann2013c} and an analysis of 23 K and early M
interferometry targets in \citet{Boyajian2012}.  A quadratic function
in $V$-$J$ color was fit to the BC values and the best-fit polynomial
was found to be ${\rm BC} = 0.69 + 0.44\left(V-J\right) -
0.037\left(V-J\right)^2$, with a scatter of only 0.03 magnitudes.
This was applied to 1068 CONCH-SHELL stars with {\it Hipparcos}
parallaxes to calculate luminosities.  We estimated masses from the
absolute $K$ magnitudes and the mass-luminosity relation of
\citet{Delfosse2000}.

\begin{figure}
\includegraphics[width=84mm]{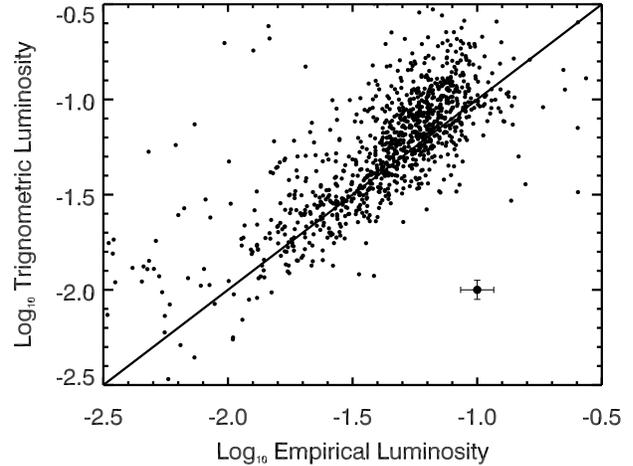}
\caption{Luminosities (solar units) based on best-fit
  \teff{} and the empirical relations in \citet{Mann2013c} compared to
  values calculated from {\it Hipparcos} parallaxes, $J$-band
  magnitudes, and a bolometric correction.  The point with error bars
  indicates the median uncertainties.}
 \label{fig.luminosity}
\end{figure}

\begin{figure}
\includegraphics[width=84mm]{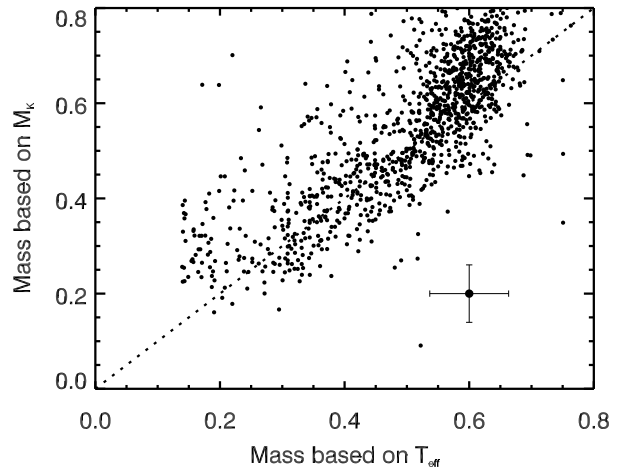}
\caption{Masses (solar units) based on best-fit \teff{} and the
  empirical relations in \citet{Mann2013c} compared to values
  calculated from {\it Hipparcos} parallaxes, $K$-band magnitudes, and
  the mass-luminosity relation of \citet{Delfosse2000}.  The point
  with error bars indicates the median uncertainties.}
 \label{fig.mass}
\end{figure}

We compare spectroscopic-based luminosities to trigonometric values,
both in solar units, in Fig. \ref{fig.luminosity}.  Figure
\ref{fig.mass} compares estimates of stellar mass. All sources of
formal error, including that from the bolometric correction, are
included.  The weighted mean difference (spectroscopic - trignometric)
between the logarithmic luminosities is $0.082 \pm 0.003$ dex.  The
average $\chi^2$ is 3.6 and 24 stars are more than 5$\sigma$ away from
the line of equality (circled points).  This is almost certainly the
result of (i) underestimation of the errors in \teff{} and the
sensitivity of our luminosity esstimates to \teff{}; (ii)
spectroscopic undersestimates of \teff{} and $L_*$ for late K stars
where there are few informative features in medium-resolution spectra.
The empirical relations of \citet{Mann2013c} are valid only for
main-sequence, inactive stars and highly active and/or very young
stars may contribute to the dispersion.  Removing stars with \ha{} in
emission (see Section \ref{sec.activity}) slightly reduces the number
of $5\sigma$ outliers and mean $\chi^2$.

We compare spectroscopic masses to those based directly on parallaxes
and $K$ magnitudes in Fig. \ref{fig.mass}.  The weighted mean
fractional difference (spectroscopic-parallax) is -6.3$\pm$0.9\% and the mean
$\chi^2$ is 0.69. Figures \ref{fig.luminosity} and \ref{fig.mass} are
not independent because the empirical curves from \citet{Boyajian2012}
and \citet{Mann2013c} are based on a mass-luminosity relation
\citep{Henry1993}.  Users of CONCH-SHELL may wish to substitute masses
and luminosities based directly on absolute $K$-magnitudes and a
mass-luminosity relation for those stars with parallaxes.

\subsection[]{Activity: H$\alpha$ Emission}
\label{sec.activity}

\begin{figure}
\includegraphics[width=84mm]{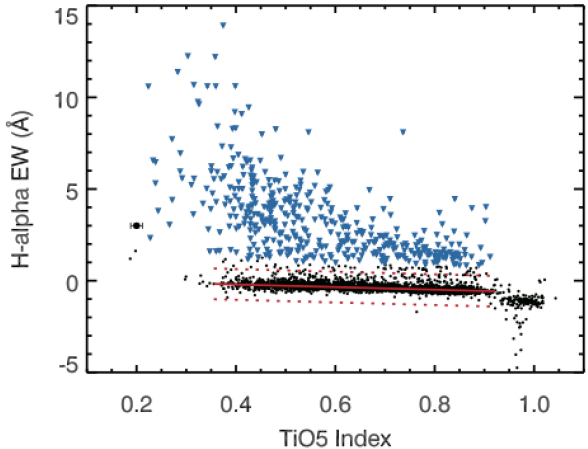}
\caption{\ha{} equivalent width vs. TiO5 index.  Blue points are
  spectra of stars with significant ($3\sigma$) emission.  A few stars
  with TiO5 $\sim 1$ (stars with late K or even earlier spectral
  types) appear to show H$\alpha$ in absorption.}
 \label{fig.halpha}
\end{figure}

The equivalent width of \ha{} was calculated by shifting each spectrum
to the rest frame using the wavelength offset produced when matching
our spectra to the PHOENIX model atmospheres (Section
\ref{sec.params}).  Following \citet{Lepine2013}, the 14\AA-wide
spectral region between 6557.61 and 6571.61\AA{} (in air) was used to
compute the EW of \ha{}, and 6500-6550\AA{} and 6575-6625\AA{} regions
were used to compute the continuum.  Errors were calculated using the
Monte Carlo method assuming Gaussian-distributed noise and random
wavelength calibration errors with an RMS of 0.5\AA{}.  Values of EW
are plotted vs. the TiO5 index in Fig. \ref{fig.halpha}.  Inactive
stars with no \ha{} emission have negative EW values because of the
spectral slope between the \ha{} region and the continuum regions.  We
fit a quadratic function with wavelegnth to these values and
calculated the intrinsic width of the locus to be 0.42\AA{} after
subtracting formal errors.  Significant ($3\sigma$) emission is seen
in 404 stars or about 13\% of stars with spectra that cover the \ha{}
region, and there is a marked increase in the envelope of EW values
towards cooler temperatures or later spectral types, as expected
\citep[e.g.,][]{Stassun2011}.

The 13\% active fraction and the trend with spectral type are
consistent with previous studies of M dwarf activity
\citep[e.g.,][]{West2008,Gizis2000}, which find typical active
fractions for early-M dwarfs (M0-M3) to be $\sim$10\%. However, this
active fraction significantly increases for later spectral subtypes,
reaching $\sim$30\% by M4 and \textgreater 80\% by M7. Moreover,
\citet{West2008} showed that the active M dwarf fraction decreases
with vertical distance $z$ from the galactic plane, e.g., for M3
dwarfs, $\sim$40\% at 25 \textless $\vert$z$\vert$\textless 50 pc
compared to $\sim$10\% at 150 \textless $\vert$z$\vert$\textless175
pc).  Our intermediate value reflects the opposing influences of the
dominance of late K and early M dwarfs over mid-M dwarfs in the
CONCH-SHELL catalog (Fig. \ref{fig.sptype}), which have a lower active
fraction, and the bias towards nearby stars which are close to the
galactic plane and have a higher active fraction.

\subsection[]{Multiplicity}

The SNIFS image cubes provide spatial information that can be used to
search for binaries. SNIFS image cubes cover $6 \times 6$~arcsec
fields of view with 0.4~arcsec pixels \citep{Aldering2002,Lantz2004}.
This limited number of pixels and low spatial resolution prohibited
the use of Gaussian source finders to identify companions.  Instead, a
principal component analysis of the two-dimensional, white-light
version of the SNIFS image cubes and by-eye checks were used to
identify binaries. The principal axes were calculated as the
eigenvectors of the spatial image moment of the background-subtracted
image.  Only pixels that were above a certain threshold multiple of
the image noise were used, where the threshold multiple scaled with
the signal-to-noise of the image.  An elongation factor $E$, the ratio
of the square root of the principle moments, and the rotation angle
$\theta$ between the principal axes and the EW-NS image coordinate
system were used as parameters to identify candidate binaries.
Criteria for $E$ and $\theta$ were set by average values from
populations of single, binary, and elongated sources identified by
eye. Candidate binaries are those with: $E > 1.16$ for any $\theta$ or
$1.03 < E < 1.16$ and $\theta < 75^{\circ}$. These complex criteria
are imposed because telescope tracking errors tend to elongate images
of point sources in the E-W direction.  This analysis was applied to
1207 SNIFS image cubes to identify 499 candidate binaries, then by-eye
inspection of the candidates confirmed 71 resolved binaries, i.e a
rate of $5.9\pm0.7$\%.

Given the spatial resolution and field of view of SNIFS, which
restricts resolvable binary separations to $\sim$1.5--4.5~arcsec, a
5.9\% binary rate is consistent with previous studies of M dwarf
multiplicity . One of the largest M dwarf multiplicity studies to date
is AstraLux \citep{Janson2012}, which included late-K to late-M
dwarfs. The AstraLux survey found 48 binaries out of 761 systems
within this separation range.  This $6.3 \pm 0.9$\% rate is perfectly
consistent with the rate we find among CONCH-SHELL stars.

\section{Comparison with the Frith et al. Catalog}
\label{sec.frith}

\citet[][F13]{Frith2013} constructed a catalog of 8479 bright ($K_S <
9$) M dwarf candidates selected from the PPMXL proper motion catalog
\citep{Roeser2010} on the basis of reduced proper motion and USNO-B
photographic {\it BRVI} and 2-MASS $JHK_S$ colors.  The F13 catalog is
most similar to CONCH-SHELL in terms of source catalog and selection
criteria, and, because all M dwarfs have $J$-$K_S>0$, we can compare
the two by imposing a $J<9$ cut on F13.  The F13 cut in $V$-$J$ color
is identical to ours ($>2.7$), although they imposed additional (but
not necessarily independent) color cuts with $B-R$, $B$-$I$, $R$-$I$,
and $I$-$J$ colors.  Their cuts in $J$-$H$ and $H$-$K_S$ colors are
not equivalent to ours but have a similar outcome, selecting stars
with $J$-$K_S$ between $\approx 0.7$ and $\approx 1$.  To separate
dwarfs from evolved stars, F13 impose a uniform reduced proper motion
criterion $H_K > 6$.  Given that M dwarfs have $J$-$K_S>0.65$, this
criterion is approximately equivalent to $H_J > 6.65$, and hence $H_V
> 6.65 + V-J$.  At $V-J = 2.7$ their cutoff in $H_V$ is about 0.5
magnitudes fainter and hence more conservative than ours.  By
$V$-$J$=5 the $H_V$ criterion of F13 is nearly three magnitudes
brighter (more relaxed) than ours, the result of F13 using a
color-independent criterion for $H_K$ and thus neglecting variation in
absolute magnitude $M_K$ along the main sequence.  However, our
selection of C-class candidates (open points in Fig. \ref{fig.redpm})
approximates their criterion because the dashed line in
Fig. \ref{fig.redpm} represents a constant $H_J = 6.15$.  Another
difference between F13 and CONCH-SHELL is that the former excluded
stars within 15 deg. of the galactic plane, and slightly farther away
at the longitude of the galactic center.

Of the 3027 F13 stars with $J<9$, 178 do not have a match in
CONCH-SHELL within 2.5 arcsec.  Of these, 48 show no detectable proper
motion ($\mu < 10$ \mpy{}) in either the Palomar plate data or the
Naval Observatory Merged Astrometric Dataset \citep{Zacharias2005}.
These may be artifacts in the PPMXL catalog.  Another 79 stars have
$\mu$ below the formal completeness limits of the SUPERBLINK catalog
(40 \mpy{} in the north, 150 \mpy{} in the south) and another 8 have
$\mu < 50$ \mpy{}.  This leaves 43 Frith stars that were missed by
SUPERBLINK: 39 are in the south.  SIMBAD searches at the locations of
the 43 reveal most to be nearby late K or M dwarf stars.

Of those F13 stars that do have SUPERBLINK matches, 306 are not in
CONCH-SHELL. Of these, 237 have revised (APASS-based) $V$-$J$ colors
that are too blue ($<2.7$).  These include many late K stars but also
some very early M-type dwarfs, the inevitable result of a catalog
selected by color rather than spectral type.  Fifty-five other stars
have $M_V$ or $H_V$ that are too bright.  Of the remaining 14 stars,
one is excluded by its parallax, two by proper motions, and 11 by
colors inconsistent with M dwarfs and their location in the ``danger
zone'' of the $H_V$ vs. $V$-$J$ diagram where extincted interlopers
may be a problem.  Among CONCH-SHELL stars, 474 are not in F13.  More
than half of these (249) are at $|b| < 15^{\circ}$ which F13 does not
cover.  There are 338 of the best candidates (class A and B) that are
not in F13.

\section{Expected Yields from Exoplanet Surveys of CONCH-SHELL Stars}
\label{sec.exoplanets}

We calculated the yield of future transit and Doppler surveys for
exoplanets around stars in the CONCH-SHELL catalog using an inference
of the planet population orbiting late-type (\teff{} $<$ 4200~K)
\kep{} stars.  Although the solar-type stars observed by \kep{} lie at
kpc distances, the few thousand M dwarfs in the target catalog are at
most a few hundred pc away and well within the galactic ``thin disk''
population \citep{Gaidos2012}.  The metallicity distribution of \kep{}
M dwarfs is also similar to that in the Solar Neighborhood
\citep{Mann2013b}.  \kep{} observations were significantly more
sensitive than the expected performance of \tess{} and the duration of
those observations more than four times longer, thus this method is
not limited by \kep{} incompleteness.  Likewise, transit surveys such
as \kep{} are generally more sensitive to small, rocky planets than
Doppler surveys because of signals in the former scale with planet
radius as $R_p^2$, while those in the latter scale as $R_p^4$
\citep{Sotin2007}.  The derivation of the planet population and its
distribution with radius and orbital period are presented in the
Appendix.  Briefly, we found that M dwarfs host an average of 2
planets with radius of 0.5-6\rearth{} and orbital period $P < 180$~d.
The distribution with radius peaks at $\sim 0.8$\rearth{} and the
distribution with orbital period follows a power-law with index 0.66
(Figs. \ref{fig.kepler_radius}-\ref{fig.kepler_period}).

\subsection{Transiting Exoplanet Survey Satellite}
\label{sec.tess}

We predicted detections of planets around CONCH-SHELL stars by the
Transiting Exoplanet Survey (\tess{}) mission \citep{Ricker2010}.  To
simulate the potential yield of \tess{} observations, the entire
synthetic \kep{} population was placed around 1000 Monte Carlo
replicates of each CONCH-SHELL star in which \teff{}, $M_*$, $R_*$,
and $L_*$ were drawn from Gaussian distributions with the standard
deviations as calculated in Section \ref{sec.params}.  For each
planet, the probability of a transiting orbit was calculated assuming
isotropically-distributed inclinations, Rayleigh-distributed
eccentricities (mean of 0.2), and uniformly distributed argument of
periastron.

The number of transits observed by \tess{} was drawn from a Poisson
distribution with a mean of $T/P$, where $T$ is the observation time.
The observation time was found by reconstructing \tess{} sky coverage
based on Fig. 7 in \citet{Ricker2014}.  The reconstruction consists of
104 pointings each covering $24 \times 24$~deg with dwell times of
27~days.  The pointings are evenly and symmetrically distributed
between the northern and southern ecliptic hemispheres in 26 pairs of
4 pointings each at 13 ecliptic longitudes spaced uniformly with
ecliptic longitude starting at 26.6$^{\circ}$ and ecliptic latitudes
(north or south) of 18, 42, 66, and 90$^{\circ}$.

Transit durations were calculated using the distribution of
dimensionless duration values $\Delta$ described in the Appendix.  The
noise during a single transit observation was calculated assuming a
pure photon noise contribution of 190~ppm for an $i=11$ star over 1~hr
plus 60~ppm of fixed systematic noise, added in quadrature.  The
detection threshold was set to SNR $>12$, a level where the \kep{}
false-positive rate is very low \citep{Fressin2013}.  For detected
planets we also calculated the orbit-averaged stellar irradiation as
in terrestrial units $L_*/4\pi a^2$, ignoring the small effect of a
non-zero eccentricity.

We calculated the fraction of planets detected by averaging over all
Monte Carlo replicates of each star and multiplying by the total
occurrence $f = 2$.  We calculated the stellar irradiation in
terrestrial units using $S = L_* (P/365.24\,{\rm d})^{-4/3}
M_*^{-2/3}$ and we ascertained whether planets orbit in the habitable
zone described by the \teff{}-dependent ``runaway'' and maximum CO$_2$
greenhouse limits on $S$ proscribed in \citet{Kopparapu2013}.

\begin{figure}
\includegraphics[width=84mm]{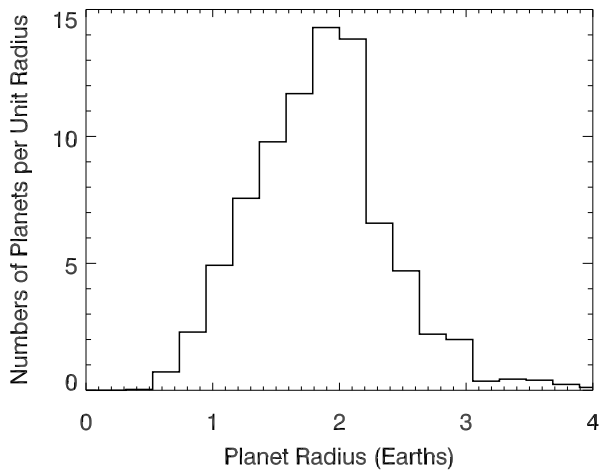}
\caption{Predicted distribution of radius of planets detected at SNR $>12$ by {\it
    TESS} around CONCH-SHELL stars.  The total number of expected
  detections is $\sim$17.}
 \label{fig.tess_radius}
\end{figure}

We estimate that TESS will observe 87\% of CONCH-SHELL stars and
that it will find $\sim 17$ planets, with only a 1.3\% chance of
finding a planet in the habitable zone of one of these stars.  If the
detection threshold is relaxed to SNR$>7.1$, the predicted number of
detections rises to 26.6, but at the expense of an elevated chance of
including false positives.  Figure \ref{fig.tess_radius} show the
distribution of predicted TESS discoveries with planet radius,
peaking at 2\rearth{} and falling precipitously by 1\rearth.  We find
that the star most likely to have a detectable planet is
PM~I19074+5905 (LSPM J1907+5905), at 2.2\%.  It is a mid-M type star
with a comparatively small radius located close to the ecliptic pole
where observations by TESS will be nearly continuous.

TESS detections of planet around CONCH-SHELL stars, especially planets
in habitable zones, is limited by the short observing intervals and
biased toward short-period orbits, where, according to \kep{}
statistics, there are fewer planets (Fig. \ref{fig.kepler_period}).
It is also limited by higher photometric noise compared to \kep{} and
the rapid decline in M dwarf planet population with increasing radius
(Fig. \ref{fig.kepler_radius}).  The distribution of simulated
detections with the \teff{} of the host star increases with cooler
\teff{}, peaking at $\sim 3500$K.  Cooler stars have smaller radii and
planets produce larger transit depths, but they also tend to be
fainter, and observations have higher noise.  Assuming the M dwarf
planet population does not depend on host star mass, the balance
between these trends in ideal surveys favors lower \teff{}.  Below
3500K, predicted TESS detections fall; some of this is due to the
\teff{} distribution of the magnitude-limited CONCH-SHELL catalog
itself (Fig. \ref{fig.teff}).  However, the distribution of detections
{\it per star} also turns over at about 3300K, or spectral
types M3-M4, suggesting that the pursuit of even cooler stars may not
be very profitable.  The $i$-magnitude limit of CONCH-SHELL is about
10.5 at the K-M dwarf boundary, thus there are additional, fainter
stars with these spectral types which could be included, but of course
these will be less attractive targets for follow-up.  An extended TESS
mission consisting of a single set of four $24 \times 24$~deg. fields
will detect many more systems per star, but for fewer stars.  A few of
the largest planets might even be detectable by ground-based surveys
such as MEarth \citep{Berta2012}.  

\subsection{Infrared Doppler Radial Velocity Survey}
\label{sec.doppler}

We simulated the yield of a hypothetical Doppler radial velocity
survey of a subset of these M dwarfs, such as those proposed for the
CARMENES \citep{Quirrenbach2012}, Habitable Planet Finder
\citep{Mahadevan2012}, IRD \citep{Tamura2012}, or SPIRou
\citep{Thibault2012} infrared spectrographs.  We assumed the following
survey parameters: (i) 300 survey nights over 5 years; (ii) 10 minute
integration per measurement plus two minutes overhead and calibration
and thus observations of 50 stars per night, or 15,000 observations.
We assumed combined measurement error and photosphere ``jitter'' of
2~m~sec$^{-1}$.  These parameters are similar to proposed surveys but
we do not attempt to replicate any specific survey.  Extensive
numerical simulations have shown that for systems with a single
dominant planet, $\sim$11 radial velocity measurements are sufficient
to identify a Keplerian signal with high confidence and distinguish it
from stellar ``jitter'' (false positive probability of $<1$\%)
\citep{Fischer2012}.  At the level of Earth masses many stars may host
multiple planets (e.g. $f = 2$ above) and disambiguation of the
measurements into separate signals requires many more measurements, of
order $\ge$50 \citep[e.g.][]{Fischer2012}.  Thus we assume 50
measurements on each of 300 stars.

We used the same planet population constructed from the \kep{} sample
described in Section \ref{sec.tess}.  To translate planet radii into
masses, we employed two different mass-radius relations.  While the
mass-radius relations for rocky planets is roughly $M_p \sim R_p^4$,
there is tentative evidence that planets larger than $\sim
1.5$\rearth{} have thick gas envelopes that contribute significantly
to their radii \citep{Marcy2014,Hadden2013}.  The two relations are
$M_p \sim R_P^2$, a scaling which connects Earth, Neptune, and Saturn,
and implies increasing gas content with planet radius/mass
\citep{Lissauer2011}, or $M_P \sim R_P$ as found by \citet{Weiss2014}
over the range 1.-5-4\rearth{}.  The actual planet population around M
dwarfs undoubtedly consists of a mix of objects
\citep{Gaidos2012,Wolfgang2012} that cannot be represented by a single
mass-radius relation.  We use the two cases of $M_p \sim
R_P^{\gamma}$, where $\gamma = 1$ and $\gamma = 2$, to bracket the
possible planet yields of Doppler surveys.

\begin{figure}
\includegraphics[width=84mm]{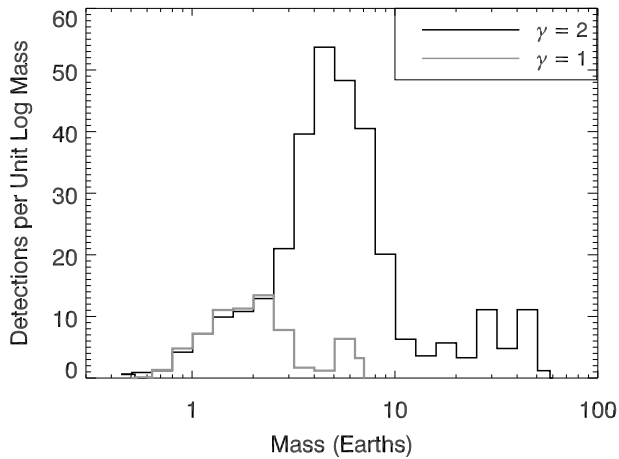}
\caption{Predicted mean mass distribution of planets detected around
  300 CONCH-SHELL stars by a hypothetical infrared Doppler radial
  velocity survey.  The planet population is that inferred around
  \kep{} M dwarfs and two different mass-radius relations of the form
  $M_P \sim R_P^{\gamma}$ are used.  The total numbers of detections
  are 7 ($\gamma = 1$) and 32 ($\gamma = 2$).}
 \label{fig.doppler}
\end{figure}

We constructed a $1000 \times$ Monte Carlo representation of the
CONCH-SHELL catalog and placed 50,000 randomly drawn planets around an
equal number of randomly drawn Monte Carlo stars.  Orbital
inclinations, eccentricities, and phases were drawn from isotropic,
Rayleigh (mean of 0.2), and uniform distributions, respectively.  We
assumed Gaussian-distributed per measurement error with RMS of 2
\mps{}.  Our detection criterion is minimalist: power in a periodogram
close to the true period with a false alarm probability $p < 0.01$,
calculated using an implementation of the method of
\citet{Scargle1982} by \citet{Horne1986}.  The dependence on
mass-radius relation is substantial: the ``heavy'' mass-radius
relation ($\gamma = 2$) leads to a prediction of 32 detections with a
peak at around 5\mearth{} (Fig. \ref{fig.doppler}) while the ``light''
relation ($\gamma = 1$) predicts only 7 detections with a peak at
$\sim 2$\mearth{}.  The predicted numbers of detections in the
habitable zone between the runaway and maximum CO$_2$ greenhouse
conditions, are 3.5 and 0.4, respectively.

These contrasting outcomes demonstrate the sensitivity of such
predictions to the mass distribution of small planets about which,
unlike the radius distribution, we know very little.  On a more
positive note, combining \kep{} transit and Doppler radial velocity
data on separate samples is one method of investigating the
mass-radius of small planets \citep{Gaidos2012,Wolfgang2012}, at least
to the extent that the populations around the two sets of stars are
statistically the same.  Of course, Doppler observations of the
transiting planets discovered by \tess{} should prove a more direct
way of investigating the nature of these worlds.

\section{Summary and Discussion}
\label{sec.summary}

We have constructed a catalog of 2970 of the brightest late K and
early M dwarfs on the sky.  These stars were selected on the basis of
parallaxes or proper motions, and visible and infrared colors.  They
will be among the most suitable targets for searches for Earth- to
Neptune-size planets by future space photometry missions and
ground-based infrared Doppler surveys.  Importantly, the bright host
stars of such planets will be amenable to follow-up observations,
e.g. spectroscopy during transits and secondary eclipses by the {\it
  James Webb} Space Telescope and a future generation of extremely
large ground-based telescopes.  We provide three data products with
this manuscript: a minimal catalog of essential stellar parameters
established for all stars and useful for selecting stars for follow-up
and exoplanet surveys (Table \ref{tab.catalog}); a full catalog with
all available parameters and their uncertainties (online
machine-readable Table 3); and a data structure containing a spectra
for each CONCH-SHELL star that was observed.

To select very cool dwarfs and screen giants and hotter stars we
applied four sets of criteria (A-D, in decreasing order of rigor).  We
obtained spectra of about 86\% of the catalog which we used to
eliminate 44 evolved or hotter stars.  We estimate that the rate of
contamination in the unscreened part of the catalog is 0.23\%, although
this rate may be higher among ``D-class'' stars.  We determined the
metallicity of 1250 stars with spectra and find a mean of [Fe/H] of
-0.07, similar to previous estimates for M dwarfs in the solar
neighborhood.  For about 13\% of the stars the Balmer \ha{} line is
seen in emission and there is an increase in both occurrence and
equivalent width for later, cooler stars.  In addition to assigning
spectral types, we fit PHOENIX BT-SETTL model spectra to determine
effective temperatures and use empirical relations to estimate stellar
radii, luminosities, and masses.

We estimated the number of planets that should be discovered around
these stars by the NASA TESS mission.  We based our calculations on
the planet population inferred to orbit \kep{} M dwarfs.  We estimate
that about 17 planets will be detected at SNR $>12$.  The number grows
to 26 if the SNR criterion is relaxed to 7.1 (the nominal detection
threshhold of \kep{}).  The radius distribution peaks at
$\sim$2\rearth{} and only 1-2 Earth-size planets are expected.
Assuming the planet population is uniform with respect to \teff{},
most planets will be found around stars with \teff{}$\sim3500$K
(spectral type M2).  We also estimated that an infrared Doppler survey
of 300 of these stars over 300 nights will discover between 7 and 32
planets, depending on the mass-radius relation for planets smaller
than Neptune.  The expected yield of planets in circumstellar
habitable zones is 0.5-3.5 from the Doppler survey, but essentially
none from TESS, a consequence of the stronger bias of transit surveys
towards short-period orbits.

Because of selection based on proper-motion and parallaxes, our
catalog is not complete to $J < 9$.  Kinematic bias and completeness
were considered for the northern sky in \citet{Lepine2013}, who
estimated that about 95\% of M dwarfs to $J<9$ were captured by the
SUPERBLINK catalog.  The southern proper motion completeness limit is
considerably higher ($\sim 150$ \mpy{}) and thus is expected to be
less complete.  We revisited this calculation using the transverse
velocity distribution of {\it Hipparcos} stars within 100~pc
\citep{vanLeeuwen2007}, the $J$-band luminosity function of
\citet{Cruz2007} and considering stars with $M_J > 5$
\citep{Lepine2013}.  We find the completeness in the northern sky to
be 98.6\%, that in the souther sky to be 88.4\%, and the
coverage-weighted kinematic completeness for the survey to be 95.2\%.
We estimate that about 152 M dwarfs were missed due to kinematic
incompleteness of the SUPERBLINK catalog.  This figure is similar to
the 130 M dwarf candidates selected by \citet{Frith2013} from the
PPMXL catalog that exhibit detectable proper motion but were missed by
in the SUPERBLINK input catalog.  Two other sources of SUPERBLINK
incompleteness arise from saturation of the source photographic plates
in the proximity of very bright stars, as well as saturation of the
cores of stars of interest, which hinder accurate astrometry
\citep{Lepine2013}.  Our catalog is also constructed based on $V$-$J$
color rather than spectral type, and some M0 stars with blue colors,
e.g. metal-poor stars, are omitted \citep{Lepine2011}.  As an
experiment, we removed the $V$-$J > 2.7$ color criterion, but imposed
the requirement that the absolute $J$ magnitude $M_J > 5.46$, the
value of the best-fit main sequence locus at $V$-$J$=2.7.  This added
342 stars, presumably a mixture of late K and M0 spectral types, to
the class A sample, an augmentation of nearly 18\%.

We did not obtain spectra of 412 CONCH-SHELL stars and we encourage
community involvement to complete the spectroscopic survey.  The AAVSO
expects to release two more versions of the APASS photometric catalog:
DR8 will improve photometry in the northern sky, and DR9 will
re-analyze the entire catalog (A. Henden, pers. comm.). Refined
photometry can be used for improved selection of M dwarfs as well as
to flux-calibrate existing spectra.  The {\it Gaia} satellite,
launched in December 2013, will obtain parallaxes with a precisions of
about 10~$\mu$as \citep{deBruijne2012} thus allowing extremely precise
determination of distance modulus.  Measurements of bolometric flux
and effective temperature could be combined to determine radii.  These
stars can also serve as a source catalog for studies other than
exoplanets, e.g. the ultraviolet (UV) emission from active,
potentially young M dwarfs and the UV luminosity function (Ansdell et
al. in prep.)

\section*{Acknowledgments}

EG acknowledges support from NASA grants NNX10AQ36G (Astrobiology:
Exobiology \& Evolutionary Biology) and NNX11AC33G (Origins of Solar
Sytems).  We thank the dedicated staff of the MDM, South African
Astronomical, and UH88 Observatories for their support.  We thank Greg
Aldering of the Nearby SuperNova Factory project for years of
assistance and help with SNIFS.  We thank Mat\'{i}as Flores, Mar\'{i}a
Luisa Luoni, Pablo Valenzuela and Emiliano Jofr\'{e} for help with the
CASLEO spectra.  This paper uses observations made at the South
African Astronomical Observatory (SAAO).  The Complejo Astron\'{o}mico
El Leoncito (CASLEO) is operated under agreement between the Consejo
Nacional de Investigaciones Cient\'{i}ficas y T\'ecnicas de la
Rep\'ublica Argentina and the National Universities of La Plata,
C\'ordoba and San Juan.  This research has made use of NASA's
Astrophysics Data System, and the SIMBAD database and the Vizier
catalogue access tool, operated at CDS, Strasbourg, France.  It was
made possible through the use of the AAVSO Photometric All-Sky Survey
(APASS), funded by the Robert Martin Ayers Sciences Fund.  It has also
made use of the NASA Exoplanet Archive, which is operated by the
California Institute of Technology, under contract with the National
Aeronautics and Space Administration under the Exoplanet Exploration
Program.  Lastly, we thank an anonymous referee for a rapid and
thorough review of an earlier version of this manuscript.

\begin{table*}
\centering
\begin{minipage}{140mm}
\caption{Telescopes \& Instruments Used to Obtain Spectra \label{tab.telescopes}}
\begin{tabular}{@{}lrrlrrl@{}}
\hline
\multicolumn{1}{c}{Telescope} & \multicolumn{1}{c}{Latitude} & \multicolumn{1}{c}{Longitude} & \multicolumn{1}{c}{Instrument} & \multicolumn{1}{c}{FHWM (\AA)} & \multicolumn{1}{c}{Spectra} & \multicolumn{1}{c}{UT Dates}\\
\hline
     MDM/McGraw-Hill & 31.95173~N & 111.61664~W &   Mark~III &  5.4 & 1098 &  2002/06/22- 2013/10/25 \\
     MDM/McGraw-Hill & 31.95173~N & 111.61664~W &       CCDS &  5.4 &  113 &  2011/06/17- 2013/10/14 \\
         MKO/UH~2.2m & 19.82303~N & 155.46937~W &      SNIFS &  6.5 & 1338 &  2009/02/22- 2014/02/24 \\
 SAAO/Radcliffe~1.9m & 32.46127~S &  20.81167~E &   RC~Spec. &  7.4 &  150 &  2013/09/18- 2013/09/25 \\
 CASLEO/Sahade~2.15m & 31.79917~S &  69.30333~W &      REOSC & 13.9 &  372 &  2010/09/11- 2013/10/29 \\
\hline
\end{tabular}
\end{minipage}
\end{table*}

\begin{table*}
\centering
\begin{minipage}{140mm}
\caption{CONCH-SHELL Catalog (Minimal) \label{tab.catalog}}
\begin{tabular}{@{}lrrrrrrrrrrrrl@{}}
\hline
\multicolumn{1}{c}{SUPERBLINK} & \multicolumn{1}{c}{RA (J2000)} & \multicolumn{1}{c}{Dec (J2000)} & \multicolumn{1}{c}{V} & \multicolumn{1}{c}{J} & \multicolumn{1}{c}{d} & \multicolumn{1}{c}{Class\footnote{See manuscript for selection criteria.}} & \multicolumn{1}{c}{SpT} & \multicolumn{1}{c}{[Fe/H]} & \multicolumn{1}{c}{$T_{\rm eff}$} & \multicolumn{1}{c}{$R_*$} & \multicolumn{1}{c}{$L_*$} & \multicolumn{1}{c}{$M_*$} \\
 & hh mm ss.s &  dd mm ss &  &  & \multicolumn{1}{c}{(pc)} &  &  &  & \multicolumn{1}{c}{(K)} & \multicolumn{3}{c}{(solar units\footnote{Negative values are upper limits.})} \\
\hline
PM I00005-0533   & 00 00 34.8 & -05 33 07 & 12.30 & 9.00 &  ---  & B &  M2 & -0.04 & 3755 & 0.51 & 0.045 & 0.55 \\
PM I00012+1358N  & 00 01 13.2 & +13 58 30 & 10.61 & 7.80 &  35.1 & A &  -- &  ---  & 3962 & 0.58 & 0.073 & 0.62 \\
PM I00013+3416   & 00 01 24.0 & +34 16 54 & 11.41 & 8.60 &  38.5 & A &  M0 &  ---  & 4157 & 0.63 & 0.107 & 0.67 \\
PM I00014-1656   & 00 01 25.8 & -16 56 54 & 10.85 & 8.02 &  32.1 & A &  M0 & -0.45 & 4021 & 0.60 & 0.082 & 0.64 \\
PM I00021-6816E  & 00 02 09.3 & -68 16 53 & 10.42 & 7.67 &  15.3 & A &  -- &  ---  & 4008 & 0.59 & 0.080 & 0.63 \\
PM I00033+0441   & 00 03 19.0 & +04 41 12 & 12.13 & 8.83 &  29.2 & A &  M1 & -0.13 & 3745 & 0.51 & 0.044 & 0.54 \\
PM I00046-4044   & 00 04 36.5 & -40 44 02 & 12.85 & 8.60 &  13.0 & A &  M4 &  ---  & 3351 & 0.28 & 0.010 & 0.26 \\
PM I00051+4547   & 00 05 10.9 & +45 47 11 & 10.00 & 6.70 &  11.3 & A &  M3 &  ---  & 3728 & 0.50 & 0.042 & 0.53 \\
PM I00054-3721   & 00 05 24.4 & -37 21 26 &  8.62 & 5.33 &   4.3 & A &  M2 &  ---  & 3589 & 0.44 & 0.028 & 0.46 \\
PM I00054-5002   & 00 05 25.0 & -50 02 53 & 12.05 & 8.55 &  ---  & B &  M2 &  ---  & 3571 & 0.43 & 0.026 & 0.44 \\
\hline
\end{tabular}
\end{minipage}
\end{table*}

\appendix
\section{The Planet Population around \kep{} M Dwarfs}
\label{sec.keplerpop}

We used the method of iterative simulation \citep{Gelman1992} to
compute the intrinsic distributions of planets with radius and orbital
period.  A large simulated population of planets is given initial
uniform and logarithmic distributions of radius and orbital period,
respectively.  The planets are randomly placed around simulated \kep{}
M dwarfs.  Planets that transit and are ``detected'' according to the
specified signal-to-noise criterion are replaced with {\it observed}
\kep{} planets selected randomly with replacement.  The trial planets
are re-shuffled among the simulated stars and the process repeated
until the properties of the simulated detections mimic those of the
observations.  The entire simulated population then represents the
intrinsic population.  If one trial planet is placed around each star,
the overall occurrence rate is the ratio of the observed number of
\kep{} planets to the number of simulated detections.

Stellar parameters of late-type dwarfs (\teff{} $<4200$K) observed by
\kep{} were estimated by Bayesian inference and a combination of
\kep{} Input Catalog photometry \citep{Brown2011} corrected to the
Sloan system \citep{Pinsonneault2012}, the Dartmouth Stellar Evolution
models \citep{Dotter2008} and priors on stellar metallicity, age,
mass, and distance \citep{Gaidos2013}.  The standard deviation between
these \teff{} values and those obtained by spectroscopy is $\sim
130$K.  We obtained candidate planet data from the Q1-Q16 Kepler
Object of Interest (KOI) catalog available on the NASA Exoplanet
Archive \citep{Akeson2013} and adjusted the planet radii using the
revised stellar radii.

We started the simulations with a planet around each \kep{} star with
radius drawn from a uniform distribution over 0.5-5\rearth{} and
orbital period $P$ drawn from a logarithmic distribution over
1-180~days.  We calculated a mean transit probability averaging over
eccentricity $e$ and argument of periastron $\omega$, and assuming
uniformly distributed $\omega$ and Rayleigh-distributed eccentricities
$\eta(e)$ with mean $e = 0.2$ \citep{Moorhead2011a}.  For each
transiting system we selected a transit duration $t = \Delta P^{1/3}
\tau^{2/3}$, where $\tau = \sqrt{3/\left(\pi^2G\rho_*\right)}$ (the
stellar free-fall time) and $\Delta$ is a dimensionless term that is a
function of $e$, $\omega$ and the transit impact parameter $b$
\citep{Gaidos2013}.  To draw values of $\Delta$ with the correct
statistical properties we compared values of a random uniform variable
$x \in [0,1]$ with the cumulative distribution of $\Delta$ averaged
over $e$ and $\omega$.  Since $b$ is also uniformly distributed, we
calculated orbit-averaged $\langle b \rangle$ as a function of
$\Delta$;
\begin{equation}
\bar{b} = \int \frac{d\omega}{2\pi} \int de \, \eta(e) \sqrt{1 -
  \Delta^2\frac{\left(1 + e \cos \omega \right)^2}{\left(1 - e^2\right)}}.
\end{equation}
Then for each $\bar{b} = x$ we found the corresponding $\Delta$.

The combined signal-to-noise of transits of a planet over all quarters is:
\begin{equation}
SNR = \left(\frac{R_p}{R_*}\right)^2 \sqrt{\sum_{i=1}^{16} \frac{n_i} {\sigma_i(t)^2}},
\end{equation}
where $n_i$ is the number of transits in the $i$th quarter and
$\sigma_i$ is the effective photometric noise over the transit
duration $t$.  We adopted $n_i \approx 90 P^{-1}$, where $P$ is in
days, for all quarters where a star was observed, and estimated
$\sum_{i=1}^{16} \sigma_i^{-2}$ at a given $t$ by interpolating or
extrapolating from values at 3, 6, and 12~hr estimated with the
corresponding values of the Combined Differential Photometric
Precision (CDPP) for each star \citep{Christiansen2012}.  We performed
power-law fits to the CDPP values; the distribution of power-law index
values peaks sharply at $\approx -0.8$. Since $-1$ is pure
uncorrelated (``white'') noise, this indicates that errors in \kep{}
photometry are slightly ``red'' due to correlated errors in photometry
and/or stellar variability.  We required a simulated SNR~$>12$ for
detection; at this threshold the rate of false positives among actual
\kep{} candidates is very low \citep{Fressin2013} and we neglect any
such adjustment.

To account for uncertainties in planet radius we substituted each KOI
by a set of Monte Carlo realizations with a distribution in radius
with standard deviation equal to the formal error.  Errors in
transiting planet radius are usually dominated by errors in stellar
radius, which for most \kep{} stars are governed by the degeneracy
between possible stellar parameters and photometric colors.  Although
these errors can be very large and non-Gaussian for F- and G-type
stars, the errors for M-type stars are better behaved because of the
large separation between the main sequence and giant branches in color
space \citep{Gaidos2013}.

However, the assumption that errors are Gaussian-distributed cannot
hold in the case of planets with radii near the detection limit of
\kep{}.  Such a planet is unlikely to be actually smaller than the
observed value because it would not have been detected; instead the
radius is more likely to be an underestimate of a larger value.  To
account for this detection bias, we modified the Gaussian error
distribution of each planet by a prior function which is the fraction
of stars around which the planet could be detected if it had a given
radius.  For large planets that were easily detected by \kep{}, the
error distribution remains Gaussian.  But for small planets, the error
distribution becomes significantly skewed, with a Gaussian tail at
larger radii and a truncated wing at smaller radii.  Uncertainties
were established from 100 bootstrap replicates of the KOI planet
sample (sampling with replacement), where the total sample size is
allowed to vary as a Poisson deviate.

\begin{figure}
\includegraphics[width=84mm]{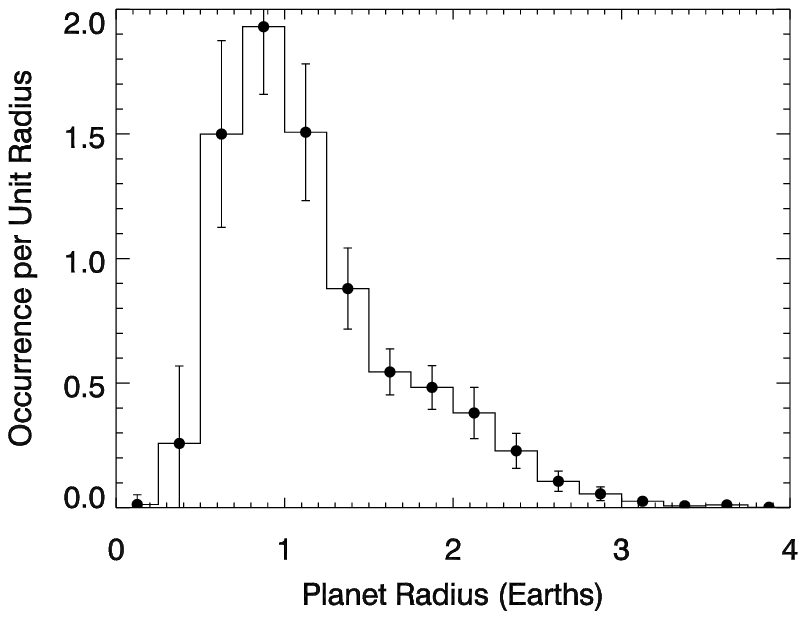}
\caption{Inferred radius distribution of planets around \kep{} M
  dwarfs with $P < 180$~days using the method of iterative simulation.
  The total occurrence is $f = 2.01 \pm 0.36$.  Uncertainties were
  established from 100 boostrap replicates.}
 \label{fig.kepler_radius}
\end{figure}

\begin{figure}
\includegraphics[width=84mm]{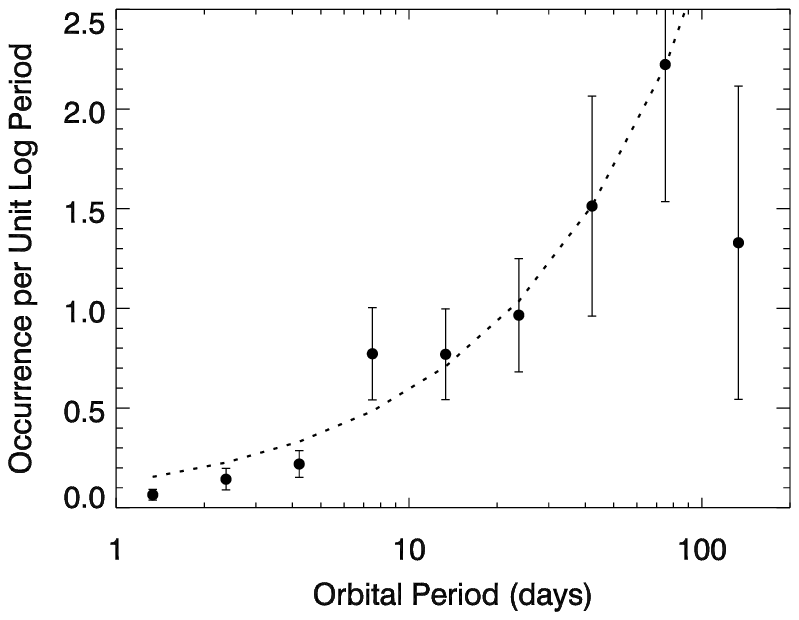}
\caption{Inferred orbital period distribution of planets with $P <
  180$~days around \kep{} M dwarfs.  Uncertainties were established from
  25 boostrap replicates.  The dashed line is a power law with
  best-fit index of 0.66.}
 \label{fig.kepler_period}
\end{figure}

The inferred radius and period distributions are shown in
Figs. \ref{fig.kepler_radius} and \ref{fig.kepler_period}.  The radius
distribution peaks near 1\rearth{} and the $\log P$ distribution
follows a power-law with index of 0.66 (determined by maximum
likelihood).  The paucity of Neptune-size (3.88\rearth) planets around
\kep{} stars is evident, and consistent with previous analyses
\citep{Howard2012,Gaidos2014}.  The total occurrence of \kep{} planets
with 0.5-6\rearth{} and $P = 1-180$~days is $f = 2.01 \pm 0.36$.  In
comparison, \citet{Swift2013} estimated $f = 1.0 \pm 0.1$ for planets
closer than $\sim 0.3$ AU, corresponding roughly to 85-day orbits.
Based on a power-law period distribution with index 0.66
(Fig. \ref{fig.kepler_period}), their result would be $f=1.64$ if
extrapolated to 180~days.  Further, the \citet{Swift2013} estimate did
not include reliable radius estimates from individual stars and
accounts only for geometric transit probability, not transit signal
detection; thus it should be regarded as a lower limit.

\label{lastpage}
\end{document}